\documentclass{article}

\usepackage{arxiv}

\usepackage[utf8]{inputenc} % allow utf-8 input
\usepackage[T1]{fontenc}    % use 8-bit T1 fonts
\usepackage{hyperref}       % hyperlinks
\usepackage{url}            % simple URL typesetting
\usepackage{booktabs}       % professional-quality tables
\usepackage{amsfonts}       % blackboard math symbols
\usepackage{nicefrac}       % compact symbols for 1/2, etc.
\usepackage{microtype}      % microtypography
\usepackage{graphicx}
\usepackage{natbib}
\usepackage{subcaption}
\usepackage{amsmath}
\usepackage{makecell}
\usepackage{xcolor}

\title{Controllable Lyrics-to-Melody Generation}

\date{} 					% Or removing it

\author{ Zhe Zhang, Yi Yu\thanks{Yi Yu is the corresponding author of this work.} , Atsuhiro Takasu \\
	Digital Content and Media Sciences Research Division\\
	National Institute of Informatics\\
	SOKENDAI \\
    Tokyo, Japan \\
	\texttt{\{zhe, yiyu, takasu\}@nii.ac.jp} \\
}

\renewcommand{\headeright}{}
\renewcommand{\undertitle}{}
\renewcommand{\shorttitle}{Controllable Lyrics-to-Melody Generation}

\begin{document}
\maketitle

\begin{abstract}
Lyrics-to-melody generation is an interesting and challenging topic in AI music research field. Due to the difficulty of learning the correlations between lyrics and melody, previous methods suffer from low generation quality and lack of controllability. Controllability of generative models enables human interaction with models to generate desired contents, which is especially important in music generation tasks towards human-centered AI that can facilitate musicians in creative activities. To address these issues, we propose a controllable lyrics-to-melody generation network, ConL2M, which is able to generate realistic melodies from lyrics in user-desired musical style. Our work contains three main novelties: 1) To model the dependencies of music attributes cross multiple sequences, inter-branch memory fusion (Memofu) is proposed to enable information flow between multi-branch stacked LSTM architecture; 2) Reference style embedding (RSE) is proposed to improve the quality of generation as well as control the musical style of generated melodies; 3) Sequence-level statistical loss (SeqLoss) is proposed to help the model learn sequence-level features of melodies given lyrics. Verified by evaluation metrics for music quality and controllability, initial study of controllable lyrics-to-melody generation shows better generation quality and the feasibility of interacting with users to generate the melodies in desired musical styles when given lyrics.
\end{abstract}

\keywords{melody generation from lyrics, controllable music generation, LSTM, GAN}

\section{Introduction}

Lyrics-to-melody generation aims to predict realistic melodies that match the meaning and rhythm of the given lyrics, similar to a human songwriter composing a song based on lyrics. Songwriting, as one of the creative human endeavors \cite{wiggins_preliminary_2006}, is a challenging task for AI to mimic.

Researchers have proposed various methods for lyrics-to-melody generation. A large paired lyrics and melody dataset was built in \cite{yu_conditional_2021}. A conditional LSTM-GAN architecture and several objective metrics are also proposed to evaluate the quality of generated melodies. In \cite{srivastava_melody_2022}, three-branch conditional LSTM-GAN are proposed to further improve generation quality. In addition to supervised learning studies, \cite{sheng_songmass_2020} performed masked sequence-to-sequence (MASS) pre-training on unpaired lyrics and melody data. Nevertheless, previous methods are uncontrollable when predicting melody and do not support user interactions, which restricts the users' creative ability and makes it hard to achieve a better melody matching the lyrics.

Controllability is an important issue in lyrics-to-melody generation tasks. Controllability of generative models enables human control of the generation process. With the steering information from human interactions, controllable models are able to generate user-desired content. In particular, controllable lyric-to-melody generation aims to generate melodies that are not only realistic and matched with the given lyrics, but also in the human-desired musical style. In music generation tasks, it is hard to generate a melody that is satisfied by all users because music generation itself is a subjective task and different users have different biases on music styles. For example, the song \textit{Yesterday} by The Beatles has more than 1,600 recorded cover versions, which share the same lyrics but vary a lot in melody contours and rhythm patterns. Such nature of music makes it necessary for music generation models to be controllable when being utilized in real scenarios as creative tools of humans. Besides, such correlations between lyrics and melody data also make it significantly different from other sequence-to-sequence generation tasks like machine translation, in which the source and target sequence have a stronger correspondence with each other. Therefore, controllable lyric-to-melody generation is not only a challenging task in machine learning research but also meaningful in real-world applications.

% In such tasks, cross-entropy (CE) loss are commonly used to train discrete-valued generation model. However, in lyrics-to-melody generation task, CE loss may be too restricted for the model to learn the unclear relationship between lyrics and melody, given the fact that the amount of paired data is also limited.

% Another important issue in lyrics-to-melody generation is one-to-many problem of lyrics-melody data. On one hand, any arbitrary word or syllable in lyrics can correspond to any arbitrary music note in the dataset, even the same lyrics sequence may correspond to different melody sequences. On the other hand, there is no "correct answer" for the generative model when predicting melodies, which can be judged in a very subjective way by the listeners. For example, the song \textit{Yesterday} by The Beatles has more than 1,600 recorded cover versions, which share the same lyrics but vary a lot in melody contours and rhythm patterns. This is significantly different from other sequence-to-sequence generation tasks like machine translation, in which the source and target sequence have a stronger correspondence with each other. In such tasks, cross entropy (CE) loss are commonly used to train discrete-valued generation model. However, in lyrics-to-melody generation task, CE loss may be too restrict for the model to learn the unclear relationship between lyrics and melody, given the fact that the amount of paired data is also limited.

To provide a flexible computer-based environment for facilitating the users to create high-quality and meaningful music, we propose a controllable lyrics-to-melody generation network in this paper with inter-branch memory fusion (Memofu) to enable information flow in multiple LSTM branches, reference style embedding (RSE) to control the style of generated melodies, and sequence-level statistical loss (SeqLoss) to better model the musical features. In the training stage, the style features are extracted from ground-truth melodies and fed into the model as reference signals along with input lyrics embedding. In the inference stage, human users can designate music style information to generate realistic while diverse melodies by modifying the input RSEs. The proposed model has three main contributions: 1) Our lyrics-to-melody generation model realizes controllability by adding the RSE module. In this way, the style of generated melodies can be easily controlled by a human user even with little music background; 2) Memofu mechanism is incorporated with multiple LSTM networks to help the model jointly learn to predict multiple related sequences; 3) SeqLoss is proposed to optimize our model by learning the sequence-level relationship between lyrics and melody. In addition, the proposed RSE and SeqLoss play a plug-in role in the architecture, which can be extended to other suitable tasks and provide controllability to the generation process.

\section{Related Work}

Music generation is a very popular topic in AI research. Various deep learning techniques and approaches for music generation are reviewed in \cite{briot_deep_2019, ji_comprehensive_2020, carnovalini_computational_2020}. In this work, we focus on tackling the difficulties of learning the correlations between lyrics-melody data pairs and enabling human control and interaction in the generation process. Thus, here we mainly discuss conditional melody generation tasks and controllable music generation.

\subsection{Lyrics-conditioned melody generation}

In recent years, we have witnessed rapid development in lyrics-to-melody generation. In \cite{choi_text-based_2016}, the authors demonstrated that text-based LSTM networks can generate chord progressions and drum tracks. Random forests are utilized in \cite{ackerman_algorithmic_2016} to model the pitch and rhythm of notes given lyrics. A lyrics-conditioned melody composition model with attention mechanism was proposed in \cite{bao_neural_2018}. Furthermore, a large paired lyrics and melody dataset was proposed in \cite{yu_conditional_2021}. Conditioned LSTM-GAN network and several objective metrics for evaluating the quality of generated melodies are also proposed in this work. To further improve generation quality, a three-branch conditional LSTM-GAN network is used in \cite{srivastava_melody_2022}. Masked sequence-to-sequence (MASS) pre-training is performed on unpaired lyrics data and melody data in \cite{sheng_songmass_2020} to construct a shared latent place of lyrics and melody. A conditional hybrid GAN model with relational memory core (RMC) technique was proposed for melody generation from lyrics in \cite{yu_conditional_2022}.

\subsection{Controllable music generation}

One challenging task in music generation is to capture and control the musical style and genre of generated content. Although deep networks act as powerful tools in generation tasks, black-box methods often lack controllability for generated music. To enable a degree of controllability, intermediate representations of music were manipulated to generate different content. In Generative Adversarial Networks (GANs) \cite{goodfellow_generative_2014}, different instances can be generated by sampling from a prior distribution \citep{dong_musegan_2017}. By manipulating the embedding of Auto-Encoders (VAEs) \citep{kingma_auto-encoding_2014}, the models can be controlled to some extent \cite{roberts_hierarchical_2019,chen_music_2020,wang_learning_2020}. However, high-level latent representation is hard for explicit and continuous control. Explicit music structure is introduced \cite{wu_popmnet_2020,dai_controllable_2021} as hierarchical embedding to model long-term musical dependency, but such methods often require professional music knowledge of users. The aforementioned works explore controllability in music generation without lyrics. TeleMelody \cite{ju_telemelody_2021} leveraged template-based methods to improve controllability in lyrics-to-melody generation, which constructs lyrics-to-melody generation as a two-stage task. An interpretable lyrics-to-melody generation system was proposed in \cite{duan_interpretable_2022}, which enables users to interact with the generation process and recreate music by selecting from recommended music attributes by the model.

\subsection{Novelty of this work}

Lyrics-to-melody generation is a discrete-valued multi-attribute sequence-to-sequence task in nature. Besides, human recognition of music is very subjective. Instead of pursuing the ``correct answer'' for generated results, we argue that the key point in lyrics-to-melody generation is to generate desired musical style fitting the input lyrics. Thus, we propose our ConL2M model to learn the correlations in the lyrics-melody data, improve generation quality, and enable human control of the generated melodies:

\begin{enumerate}
    \item Generating melodies from lyrics is a process to predict multiple sequences of music attributes from the same source input. Inspired by the coupled-layer architecture for LSTM proposed in \cite{liu_recurrent_2016} for multi-task learning (MTL), we proposed an inter-branch memory fusion (Memofu) technique to help multiple stacked LSTM \cite{graves_speech_2013} networks to learn the dependencies across the sequences. The proposed Memofu architecture not only realizes memory fusion between LSTMs like previous work but also keeps independent output layers to better predict different music attributes.
    \item Inspired by global style token (GST) \citep{wang_style_2018} in text-to-speech (TTS) task, we propose Reference Style Embedding (RSE) technique as an effective technique to generate realistic melodies from lyrics while allowing control on the style of generated melodies. Different from the aforementioned methods, we use explicit statistical features of melodies as RSE features to realize controllability, which requires no music background knowledge, yet is intuitively consistent with human perception of music. 
    \item To help the model learn meaningful style features of music sequences, a sequence-level statistical loss (SeqLoss) is proposed in this work to optimize the generative model. Different from the widely-used cross-entropy (CE) loss, which forces the model to learn a token-to-token correspondence, our proposed SeqLoss relaxes the token-level constraints and optimizes the model by sequence-level features instead, which improves the quality of generated music while preserving generation diversity. Prior knowledge from the dataset is also introduced to the model by SeqLoss to better update the parameters, taking the advantage of the Gumbel-softmax \cite{jang_categorical_2017} technique.
\end{enumerate}

\section{Approach}

In sequence-to-sequence generation tasks, deep learning models are trained to predict the alignment and correlation between source and target. However, in lyrics-to-melody generation task, the same lyrics sequence can match various melody sequences in different styles and genres. To overcome the difficulties of training a lyrics-to-melody generation model and enable control of the generated musical style, we propose a controllable lyrics-to-melody generation model called ConL2M, which is introduced in this section.

\subsection{Overview of ConL2M}

\begin{figure}[htbp]
    \centering
    \includegraphics[width=0.9\columnwidth]{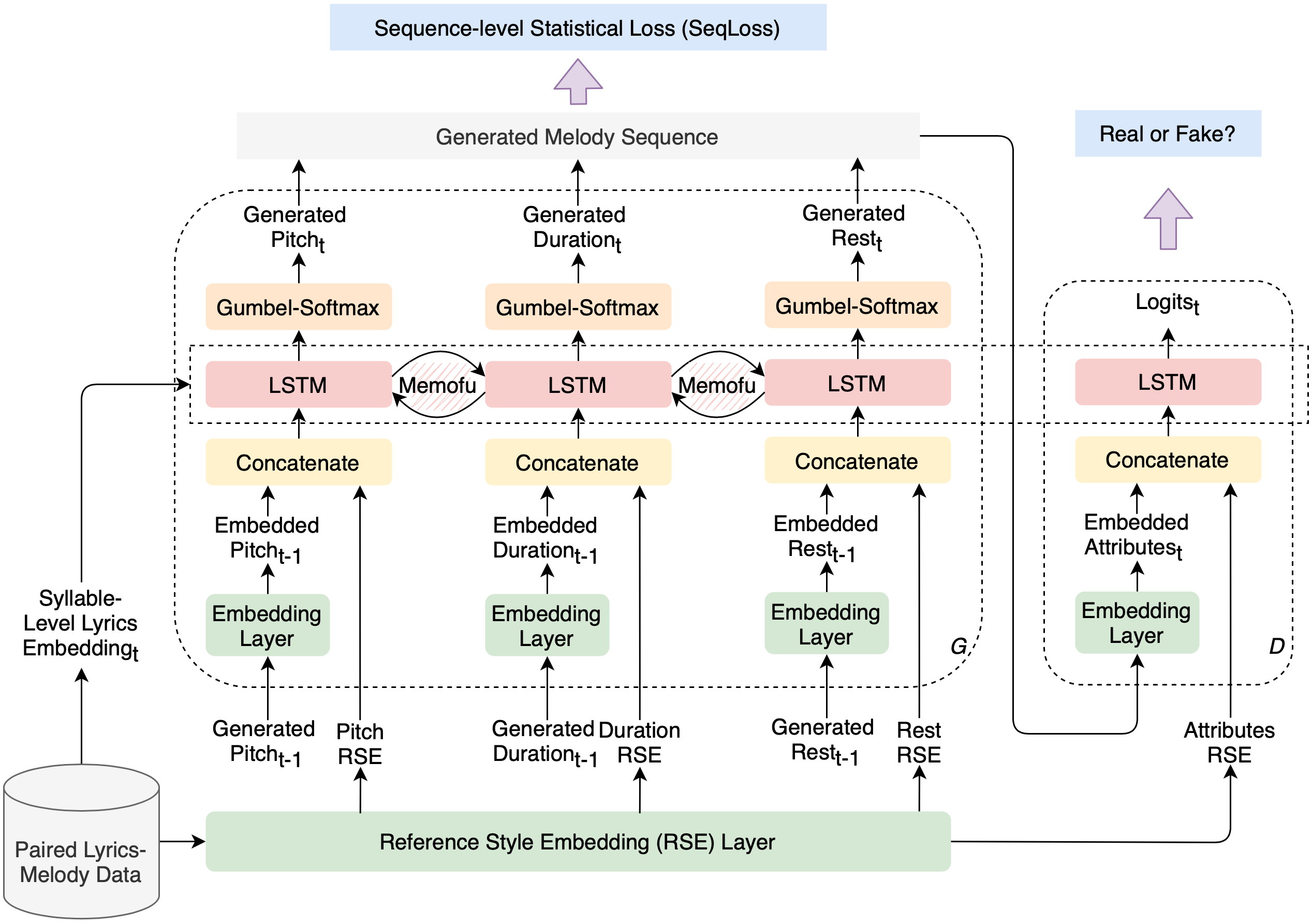}
    \caption{Overall architecture of ConL2M.}
    \label{fig:architecture}
\end{figure}

% The SeqCtrlMelody architecture is shown in Fig. \ref{fig:architecture}, which is based on Generative Adversarial Networks (GANs) \citep{goodfellow_generative_2014} with a lyrics-conditioned LSTM \citep{hochreiter_long_1997} multi-branch generator and a LSTM-based discriminator. In the generator's LSTM networks, inter-branch information fusion between different sub-networks is enabled by the proposed Memofu mechanism. In addition, RSE layer adds controllability and improves the generation quality of the model. Moreover, SeqLoss is exploited to jointly optimize the generator with Relativistic Standard GAN (RSGAN) loss \citep{jolicoeur-martineau_relativistic_2018}. 

The ConL2M architecture is shown in Fig. \ref{fig:architecture}, which is based on integrating Generative Adversarial Networks (GANs) \citep{goodfellow_generative_2014} with LSTM \citep{hochreiter_long_1997}, similar to our previous work \citep{srivastava_melody_2022} but with significant differences: i) In the generator's LSTM networks, inter-branch information fusion between different sub-networks is enabled by the proposed Memofu mechanism; ii) The proposed RSE layer adds controllability and improves the generation quality of the model; iii) The proposed SeqLoss is exploited to jointly optimize the generator with Relativistic Standard GAN (RSGAN) loss \citep{jolicoeur-martineau_relativistic_2018}.

\begin{figure}[htbp]
    \centering
    \includegraphics[width=0.9\columnwidth]{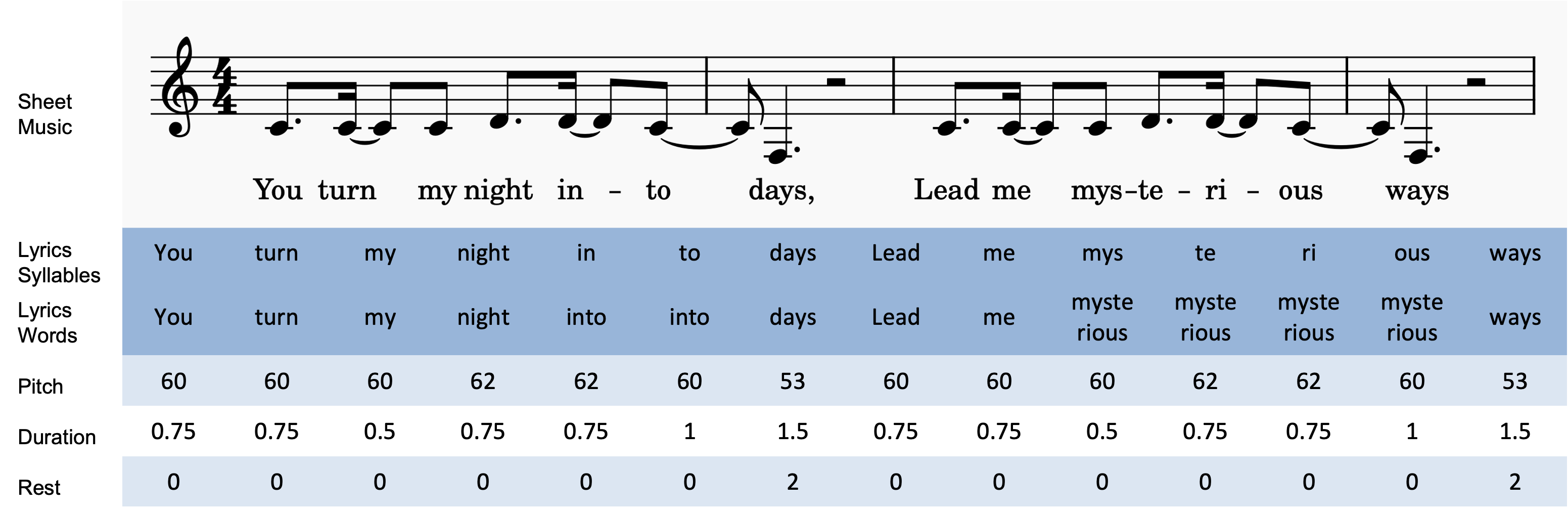}
    \caption{An example of paired lyrics-melody data.}
    \label{fig:alignment}
\end{figure}

An example of paired melody and lyrics sequence in our dataset is shown in Fig. \ref{fig:alignment}. Tokens of lyrics are embedded by pre-trained Skip-gram \citep{mikolov_distributed_2013} model in both word-level and syllable-level. The embedding vectors are then concatenated together to form the representation of lyrics token $x_t$. A note in melody is represented as a triplet of music attributes $[y^{pitch}_t, y^{duration}_t,  y^{rest}_t]$, including the note pitch in form of the corresponding MIDI number, note duration, and rest duration after the current note. The unit of note duration and rest duration is a quarter note.

In one training time step, the generated music attribute $y^{\mathrm{attr}}_{t-1}$ is concatenated with the RSE vector, which is discussed in Sec. \ref{subsec:rse}, and then fed into the corresponding LSTM branch together with the current lyrics embedding $x_t$ as the condition. By exploiting the Gumbel-softmax technique, discrete-valued sequences of music attributes are predicted by the generator. Then, model parameters are updated by optimizing both SeqLoss (discussed in Sec. \ref{subsec:seqloss}) and RSGAN loss through the discriminator.

\subsection{Inter-branch memory fusion (Memofu)} \label{subsec:memofu}

% LSTM is a kind of recurrent neural network (RNN) with the ability to model long dependencies in the sequences, which has been widely exploited in sequence generation tasks. The transition equations for a single-layer LSTM network can be defined as follows:

% \begin{align}
% \mathbf{i}_t &=\sigma\left(\mathbf{W}_i \mathbf{x}_t+\mathbf{U}_i \mathbf{h}_{t-1}+\mathbf{V}_i \mathbf{c}_{t-1}\right), \\
% \mathbf{f}_t &=\sigma\left(\mathbf{W}_f \mathbf{x}_t+\mathbf{U}_f \mathbf{h}_{t-1}+\mathbf{V}_f \mathbf{c}_{t-1}\right), \\
% \mathbf{o}_t &=\sigma\left(\mathbf{W}_o \mathbf{x}_t+\mathbf{U}_o \mathbf{h}_{t-1}+\mathbf{V}_o \mathbf{c}_t\right), \\
% \tilde{\mathbf{c}}_t &=\tanh \left(\mathbf{W}_c \mathbf{x}_t+\mathbf{U}_c \mathbf{h}_{t-1}\right), \\
% \mathbf{c}_t &=\mathbf{f}_t^i \odot \mathbf{c}_{t-1}+\mathbf{i}_t \odot \tilde{\mathbf{c}}_t, \\
% \mathbf{h}_t &=\mathbf{o}_t \odot \tanh \left(\mathbf{c}_t\right),
% \end{align}
% where $\mathbf{i}_t$, $\mathbf{f}_t$, and $\mathbf{o}_t$ denote the input gate, the forget gate, and the output gate in an LSTM unit, respectively, while $\mathbf{c}_t$ and $\mathbf{h}_t$ denote the memory cell and the hidden state. Note that the bias terms are omitted in all formulas of this paper for simplicity.

LSTM \citep{hochreiter_long_1997} is a kind of recurrent neural network (RNN) with the ability to model long dependencies in the sequences, which has been widely exploited in sequence generation tasks. At time step $t$, the transition equations for a single-layer LSTM network can be defined as follows:

\begin{align}
i_t & =\sigma\left(W_i\left[h_{t-1}, x_t\right]+b_i\right), \\
f_t & =\sigma\left(W_f\left[h_{t-1}, x_t\right]+b_f\right), \\
o_t & =\sigma\left(W_o\left[h_{t-1}, x_t\right]+b_o\right), \\
\tilde{c}_t & =\tanh \left(W_c\left[h_{t-1}, x_t\right]+b_c\right)\label{eq:lstm_c_t}, \\ 
c_t & =f_t \circ c_{t-1}+i_t \circ \tilde{c}_t\label{eq:lstm_c_t_2},  \\
h_t & =o_t \circ \tanh \left(c_t\right)\label{eq:lstm_h_t}, 
\end{align}
where $i_t$, $f_t$, and $o_t$ denote the input gate, the forget gate, and the output gate in an LSTM unit, respectively, while $c_t$ and $h_t$ denote the memory cell and the hidden state. $x_t$ is the input of the LSTM network. $W$s and $b$s are weights and bias terms. $\sigma(\cdot)$ is the sigmoid function and $\circ$ denotes the element-wise multiplication.

Based on single-layer LSTM, deep LSTM networks are proposed which can yield significant improvement over single-layer LSTM \cite{graves_speech_2013}. Deep LSTMs are created by stacking multiple LSTM layers on top of each other, with the output sequence of one layer forming the input sequence for the next. Specifically, the hidden states output by the $n$th layer can be defined as 
\begin{equation}
% \mathbf{h}_t^n=\mathcal{H}\left(\mathbf{W}_{h^{n-1} h^n} \mathbf{h}_t^{n-1}+\mathbf{W}_{h^n h^n} \mathbf{h}_{t-1}^n+\mathbf{b}_h^n\right),
% h_t^n = \mathcal{H}\left(W_{h^{n-1} h^n} h_t^{n-1} + W_{h^n h^n} h_{t-1}^n + b_h^n\right),
h_t^n = \mathcal{H}\left(W_h^{n-1 \rightarrow n} h_t^{n-1} + W_h^{n} h_{t-1}^n + b_h^n\right),
\end{equation}
where $\mathcal{H}$ denotes the activation function between layers and $h_t^{0}$ is defined as the input of LSTM $x_t$. $W_h^{n-1 \rightarrow n}$ denotes the weights of the cells from the $(n-1)$th layer to the $n$th layer and $W_h^{n}$ denotes the weights of the cells from the previous time step to the current time step in the $n$th layer.

In addition to the development of LSTMs in the depth direction, researchers also exploit multi-task learning (MTL) frameworks to jointly learn across multiple related tasks. In \cite{liu_recurrent_2016}, the authors proposed three different information sharing mechanisms to help LSTM networks learn from multiple tasks. Unlike the previous methods such as uniform-layer architecture, coupled-layer architecture, and shared-layer architecture, we proposed a memory fusion mechanism between stacked LSTM networks to model the dependencies from different sequences of musical attributes.

\begin{figure}[htbp]
    \centering
    \includegraphics[width=0.9\columnwidth]{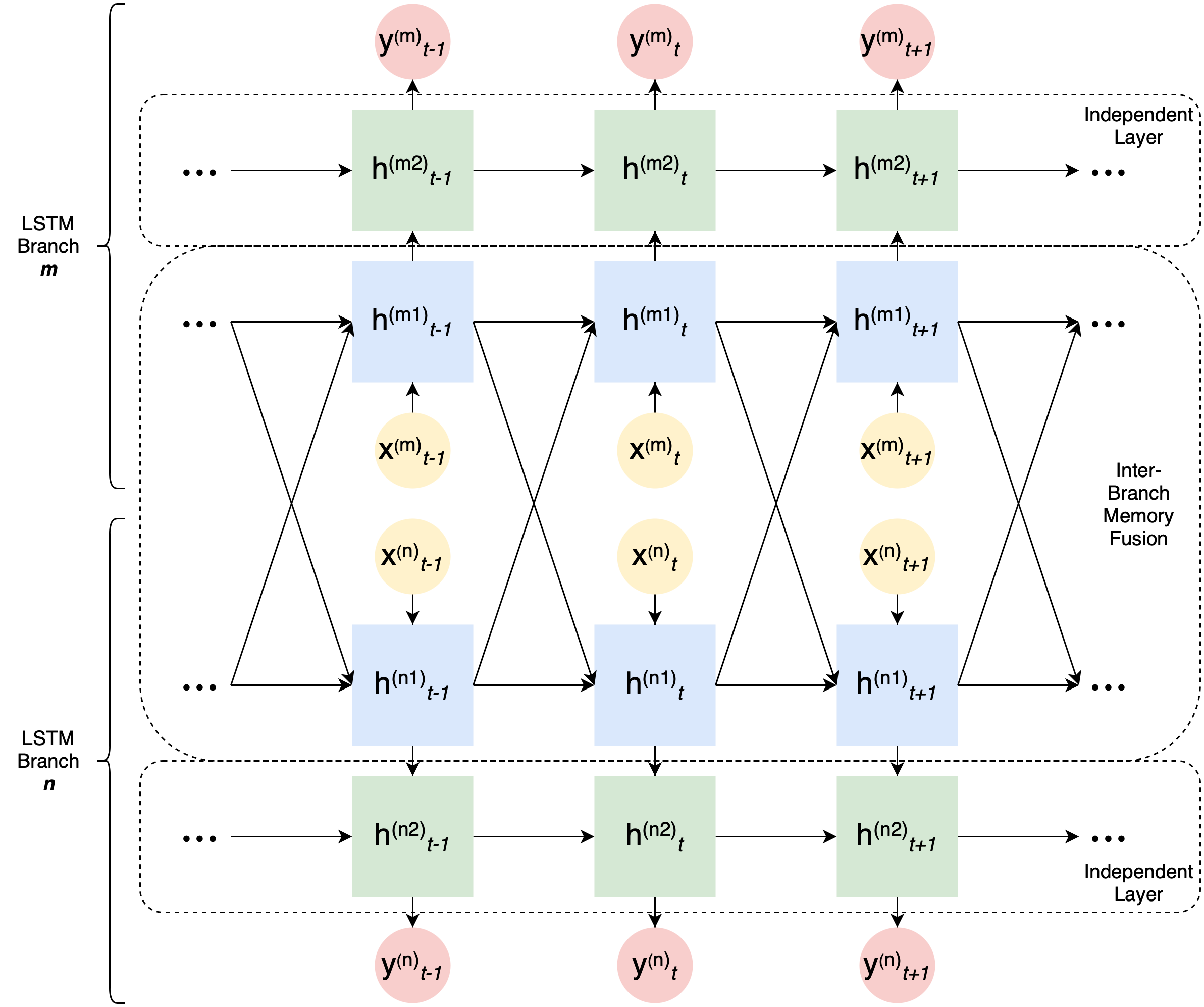}
    \caption{Illustration of two-branch stacked LSTM with memory fusion.}
    \label{fig:memofu}
\end{figure}

The architecture of the proposed LSTM architecture with inter-branch memory fusion is shown in Fig. \ref{fig:memofu}. Each branch is a two-layer stacked LSTM network, consisting of a fusion layer that connects with other branches and an independent layer that predict the output for the current attribute. 

Fig. \ref{fig:memofu} is an example of the proposed Memofu architecture of a two-branch ($m$ and $n$) setting. In our ConL2M architecture, three branches of music attribute generators fuse with each other to learn inter-branch information while the input of the three branches are the same lyrics embedding $x_t$. Specifically, the memory cells for the branch $j$ in the proposed Memofu architecture can be defined as follows:

\begin{align}
% \tilde{c}_t^{(br)_{in}} &= \tanh \left(W_c^{(br)_{in}} x_t + \sum_{i \in\{p, d, r\}} U_c^{(i \rightarrow br)} h_{t-1}^{(i)_{in}} + b_c^{(br)_{in}}\right), \label{eq:memofu_in}\\
% % \mathbf{h}_t^{(br)_{out}}=\mathcal{H}\left(\mathbf{W}_{h_{in}} \mathbf{h}_t^{{(br)_{in}}}+\mathbf{W}_{h_{out}} \mathbf{h}_{t-1}^n+\mathbf{b}_h^n\right),
% \tilde{c}_t^{(br)_{out}} &= \tanh \left(W_c^{(br)_{out}} h_t^{(br)_{in}} + U_c^{(br)_{out}} h_{t-1}^{(br)_{out}} + b_c^{(br)_{out}}\right)\label{eq:memofu_out}, 
\tilde{c}_t^{(j), \mathrm{in}} &= \tanh \left(W_{c}^{(j), \mathrm{in}} x_t + \sum_{i \in\{p, d, r\}} U_c^{(i \rightarrow j)} h_{t-1}^{(i), \mathrm{in}} + b_c^{(j), \mathrm{in}}\right), \label{eq:memofu_in}\\
% \mathbf{h}_t^{(br)_{out}}=\mathcal{H}\left(\mathbf{W}_{h_{in}} \mathbf{h}_t^{{(br)_{in}}}+\mathbf{W}_{h_{out}} \mathbf{h}_{t-1}^n+\mathbf{b}_h^n\right),
\tilde{c}_t^{(j), \mathrm{out}} &= \tanh \left(W_c^{(j), \mathrm{out}} h_t^{(j), \mathrm{in}} + U_c^{(j), \mathrm{out}} h_{t-1}^{(j), \mathrm{out}} + b_c^{(j), \mathrm{out}}\right)\label{eq:memofu_out}, 
\end{align}
where $j\in\{p, d, r\}$, and $p$, $d$, and $r$ represent the three branches in our network, corresponding to the three musical attributes, pitch, duration, and rest, respectively. The notation $\mathrm{out}$ represents the input layer as well as the fusion layer in our LSTM network (marked in blue in Fig. \ref{fig:memofu}) and $\mathrm{out}$ represents the independent output layer (marked in green in Fig. \ref{fig:memofu}). $W$s and $U$s are the weights and $b$s are bias terms.

Compared with the computation of $\tilde{c}_t$ in vanilla LSTM as shown in (\ref{eq:lstm_c_t}), the memory cells in our fusion layer not only consider the previous hidden state of the current LSTM branch, but also learn the dependencies from the previous hidden states of other LSTM branches, as shown in the $\sum$ term in (\ref{eq:memofu_in}). Meanwhile, the output layer remains independent across LSTM branches like standard stacked LSTM networks as shown in (\ref{eq:memofu_out}). Intuitively, the $in$ LSTM layer helps the model learn the inter-dependency of the different attribute sequences in the generation process while the $out$ LSTM layer focuses on predicting the distribution of the current attribute. After computing $\tilde{c}_t$s through the above memory fusion mechanism, the memory cells $c_t$s and hidden states $h_t$s are derived by (\ref{eq:lstm_c_t_2}) and (\ref{eq:lstm_h_t}). Specifically, the output of each branch is given by mapping the hidden states of the output LSTM layer $y_t = W_y h_t^{\mathrm{out}} + b_y$.

\subsection{Reference style embedding (RSE)} \label{subsec:rse}

The proposed RSE aims to help the model learn better style information of a melody sequence and make possible the controllable generation process in the inference stage with human interaction. In our lyrics-to-melody generation task, we use statistical features of target sequences of each music attribute as style features, namely range, average, and variance of the music attributes. These style features are not only straightforward to compute, but also reliable in interpreting the musical style. For example, given the same lyrics, a melody in Rock genre may have a lower duration range (stable rhythm), while a Jazz melody usually gets higher duration variance (swingy rhythm).

Detailed explanations of the meaning of the features and their relations to human perception of music are listed as follows:

\begin{itemize}
    \item Pitch Range (PR): the vocal range of the melody. Higher the PR, the melody will cover a larger vocal range. Lower the PR, the melody will fall in a narrower pitch band.
    \item Pitch Average (PA): the average absolute pitch level. Higher the PA, the melody may be sung by a singer with a higher vocal range and \textit{vise versa}.
    \item Pitch Variance (PV): the tendency of the pitch contour. Higher the PV, the melody will tend to have a more ``jumping'' feeling in its development. Lower the PV, the melody will tend to have a more stable feeling.
    \item Duration Range (DR): the degree of rhythm changes in melody. Higher the DR, the melody will have more tendency to include both long notes and short notes. Lower the DR, the melody will tend to have more similar note duration.
    \item Duration Average (DA): the average note duration level of the melody. Higher the DA, the melody will have more long notes and feel slower \textit{vise versa}.
    \item Duration Variance (DV): the complexity of the rhythm of the melody. Higher the DV, the melody tends to have more complex rhythm patterns like Hip-hop music. Lower the DV, the melody tends to have more fixed rhythm patterns like Dance music.
    \item Rest Range (RR): The longest rest note in the melody. Higher the RR, the melody will have longer rest notes within \textit{vise versa}.
    \item Rest Average (RA): The average rest note duration in the melody. Higher the RA, the melody will have more rest time \textit{vise versa}.
    \item Rest Variance (RV): The pattern of the rest notes of the melody. Higher the RV, the melody will tend to have more different rest notes \textit{vise versa}.
\end{itemize}

In practice, the above style features are calculated for each melody sequence in the data pre-processing stage. Because these statistics are taken from all the time steps of the target music attribute sequences, they contain sequence-level style information of melodies for our model to refer to. The extracted style features are continuous values (except for pitch range because pitch values are discrete), which are discretized and encoded as one-hot vectors $ref^{\mathrm{attr}}_{ft}$,
\begin{equation}
    ref^{\mathrm{attr}}_{ft} = {\mathrm{OneHot}}(ft^{\mathrm{attr}}),
\end{equation}
where $\mathrm{attr}$ denotes pitch, duration, and rest, while $ft$ denotes three features, namely range, average, and variance. Then, for each branch, the feature vectors are concatenated to form the RSE of the corresponding branch as
\begin{equation}
    \mathbf{ref}^{\mathrm{attr}} = [ref^{\mathrm{attr}}_{\mathrm{rng}}, ref^{\mathrm{attr}}_{\mathrm{avg}}, ref^{\mathrm{attr}}_{\mathrm{var}}].
\end{equation}

Then, in one time step of LSTM, RSE is concatenated with $y^{\mathrm{attr}}_{t-1}$ to form the input for each branch,
\begin{equation}
        \mathbf{input}^{\mathrm{attr}}_t = \mathrm{Concat}(y^{\mathrm{attr}}_{t-1}, \mathbf{ref}^{\mathrm{attr}}).
\end{equation}

In the recurrent training procedure, RSE acts as a sequence-level condition, being fed into LSTM in every time step of the generation process to steer the generative model. RSE is also fed into the discriminator to affect the adversarial training.

As a brief discussion, the three extracted RSE features, namely pitch, duration, and rest, are not independent of each other, i.e., the variance of a sequence can be positively correlated with the range of a sequence. Nevertheless, they still capture the different properties of data distribution in one sequence. For instance, a sequence with a large range value may only contain one data point distributed far from the mean, while a sequence with a large variance value indicates a large dispersion degree of all data points in the sequence. Our experimental results in \ref{subsec:controllability} and \ref{subsec:sheetmusic} verified that our RES technique is capable to control the musical style of the generated sequences in different aspects although there remain some correlations between the RSE inputs.

\subsection{Sequence-level statistical loss (SeqLoss)} \label{subsec:seqloss}

In discrete-valued sequence generation tasks like machine translation, CE loss often acts as a default choice for loss function. However, in lyrics-melody data, the correlation between lyrics and melody is relatively weaker and melodies in different musical styles can correspond to the same lyrics. Thus, CE loss will limit the generation diversity and generalization ability of the model, considering the fact that the existing paired lyrics-melody datasets are far from covering all the relationships and alignment between lyrics and melody in the real world. To overcome this difficulty, we propose SeqLoss to guide the model to focus on learning the sequence-level musical features.

The Gumbel-softmax relaxation technique is originally proposed to solve the non-differentiable problem of discrete-valued GAN training, which is also exploited in melody generation tasks because music attributes are discrete values in nature. Taking the advantages of Gumbel-softmax, we can derive a differentiable approximation of the label as a one-hot vector from the output logits $o$:
\begin{equation} \label{eq:gumbel}
\begin{split}
        \hat{\mathbf{p}}^{\mathrm{attr}}_{t} &= {\mathrm{GumbelSoftmax}}(o^{\mathrm{attr}}_{t-1}) \\
                            &\approx {\mathrm{OneHot}}(\arg\max(o^{\mathrm{attr}}_{t-1})),
\end{split}
\end{equation}
where $\hat{\mathbf{p}}^{\mathrm{attr}}_{t} \in \mathbb{R}^{K}$ denotes the predicted probabilities of the generated music attributes, and $K$ denotes the dimension of the embedded music attributes.

Then, we can compute the mean and variance of the sequence of generated music attributes as
\begin{gather}
    % \hat{m}^{\mathrm{attr}} = \frac{1}{T}\sum_{t=1}^{T}\sum_{k=1}^{K} k \cdot \hat{p}^{\mathrm{attr}}_{t,k},\\
    % \hat{v}^{\mathrm{attr}} = \frac{1}{T-1}\sum_{t=1}^{T}\sum_{k=1}^{K} \left(k \cdot \hat{p}^{\mathrm{attr}}_{t,k} - \hat{m}^{\mathrm{attr}}\right)^{2},
    \hat{m}^{\mathrm{attr}} = \frac{1}{T}\sum_{t=1}^{T}\left(\sum_{k=1}^{K} k \cdot \hat{p}^{\mathrm{attr}, t}_{k}\right),\\
    \hat{v}^{\mathrm{attr}} = \frac{1}{T-1}\sum_{t=1}^{T} \left(\sum_{k=1}^{K}k \cdot \hat{p}^{\mathrm{attr}, t}_{k} - \hat{m}^{\mathrm{attr}}\right)^{2},
\end{gather}
where $k\in\{1,2,...,K\}$ denotes the class labels of the attribute and $\hat{p}^{\mathrm{attr}, t}_{k}$ denotes the predicted probability of the $k$th class of the attribute in time step $t$. $T$ is the length of the sequence. Here, prior music knowledge is involved in the optimizing process to guide the model instead of only considering the difference between the probability token distributions of predictions and ground-truth.

Then, our proposed SeqLoss can be defined as
\begin{equation}
\begin{split}
    L^{\mathrm{attr}}_{\mathrm{seq}} &= \alpha_1 \cdot \mathrm{MSE}(\hat{m}^{\mathrm{attr}}, m^{\mathrm{attr}}) \\
                            &+ \alpha_2 \cdot \mathrm{MSE}(\hat{v}^{\mathrm{attr}}, v^{\mathrm{attr}}),
\end{split}
\end{equation}
where $\alpha_1$ and $\alpha_2$ are hyperparameters to weight the loss terms. $\mathrm{MSE}(\cdot,\cdot)$ denotes the computation mean square error (MSE) of two vectors of length $L$ by $\mathrm{MSE}(x,y) = \sum_{i=1}^{L}(x_i-y_i)^2$. In our implementation, MSEs of the mean and variance of the sequence of generated
music attributes are computed on every batch. Compared with traditional CE loss, SeqLoss not only improves the diverse generation ability of the model, but also introduces prior knowledge of labels to measure the distance between different label classes to optimize the model better on the specific task.

\section{Dataset}

\begin{figure}[htbp]
    \centering
    \includegraphics[width=0.9\columnwidth]{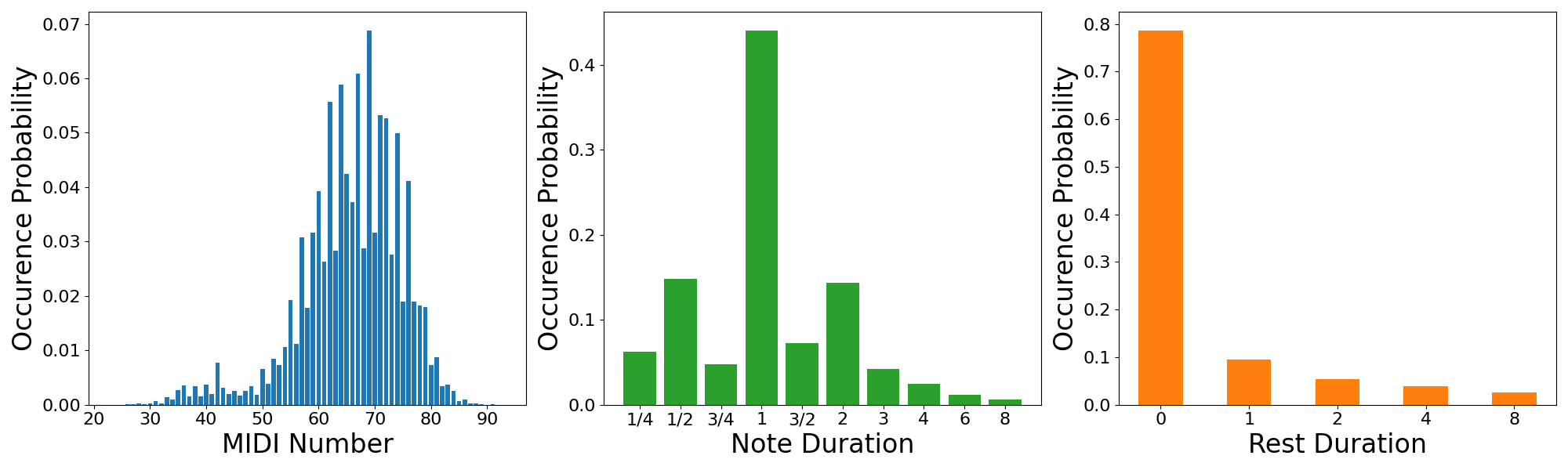}
    \caption{Distribution of music attributes in the dataset.}
    \label{fig:dataset_disttribution}
\end{figure}

In our experiment, a large paired lyrics-melody dataset proposed in \citep{yu_conditional_2021} is used to train and evaluate the models, which contains 13,251 paired lyrics-melody sequences. Each sequence pair consists of 20 syllables of lyrics and corresponding 20 notes of melodies in form of triplets of music attributes $[y^{pitch}_t, y^{duration}_t,  y^{rest}_t]$, as shown in Fig. \ref{fig:alignment}. Because some melodies in the original dataset contain unreal pitch and note duration for human singing, we filtered the dataset to keep the meaningful lyrics-melody data pairs. The details of filtering rules are shown in Table \ref{tab:filter_rules}, which is decided according to the common singing ability of human. The distribution of the music attributes after the filtering pipeline is shown in Fig. \ref{fig:dataset_disttribution}.

\begin{table}[htbp]
    \centering
    \begin{tabular}{lcc}
    \toprule
    Features & Range & Musical Meaning \\
    \midrule
    Pitch Range & 0-48 & 4 octaves \\
    Pitch Average & 36-84 & C2 to C6 \\
    Duration Range & 0-8 & 2 bars \\
    Duration Average & 0-4 & 1 bar \\
    Rest Range & 0-8 & 2 bars \\
    \bottomrule
    \end{tabular}
    \caption{Conditions for filtering data.}
    \label{tab:filter_rules}
\end{table}

After data filtering, 10,768 pairs of lyrics-melody sequences are kept to split the train, validation, and test dataset in an 8:1:1 ratio. The same train/validation/test dataset is used for training and evaluating all models.

\section{Experiment}

In this section, we discuss the experiments and evaluation results to demonstrate the generation ability and controllability of the proposed ConL2M. Objective evaluation metrics and music features are used to evaluate generation quality. Self-BLEU scores \cite{zhu_texygen_2018} are used to analyze generation diversity. Experiments to demonstrate controllability are designed and conducted. Moreover, subjective evaluation metrics for controllability are proposed in this section.

In addition, some strongest existing works of melody generation from lyrics such as TBC-LSTM-GAN \cite{srivastava_melody_2022} and C-Hybrid-GAN \cite{yu_conditional_2022} are selected as the baselines in objective evaluation and subjective evaluation, which outperform the earliest methods such as conditional LSTM-GAN and MLE \cite{yu_conditional_2021}. TBC-LSTM-GAN consists of three independent LSTM networks as melody the generator and one LSTM network as the discriminator. C-Hybrid-GAN exploits RMC technique to model the dependency inside each sequence of attributes during the training of the generator and discriminator. In addition, three competitive ablation models are implemented to compare with the proposed ConL2M:

\begin{enumerate}
    \item Memofu: multi-branch LSTM generator with the Memofu mechanism, LSTM discriminator, and RSGAN loss with Gumbel-softmax technique.
    \item SeqMemofu: multi-branch LSTM generator with Memofu mechanism, LSTM discriminator, and RSGAN loss with Gumbel-softmax technique, and jointly using the Seqloss.
    \item ConMemofu: multi-branch LSTM generator with Memofu mechanism, LSTM discriminator, and RSGAN loss with Gumbel-softmax technique, adding the RSE module.
\end{enumerate}

\subsection{Experiment setup}

To verify the effects of the proposed techniques, we trained the proposed model, ablation models, and the baseline models with the shared configurations in Table \ref{tab:config}. The generator is pre-trained for 40 epochs with CE loss. Then the generator and the discriminator are together trained with RSGAN loss for 120 epochs. Adam optimizers \cite{kingma_adam_2017} are used with $\beta_1=0.9$ and $\beta_2=0.99$. Gradient clipping of 5 is applied. The maximum temperature of Gumbel-softmax is 1,000 and the batch size is 512.

\begin{table}[h]
  \caption{Configuration of the generator and discriminator}
  \label{tab:config}
  \centering
      \begin{tabular}{ll|ccccc}
        \toprule
        \multicolumn{2}{c|}{Module} & \makecell[c]{Embedding\\Dimension} & \makecell[c]{Hidden\\Dimension} & \makecell[c]{LSTM\\Units} & \makecell[c]{Output\\Dimension}\\
        \midrule
                    & Pitch     & 128 & 32 & 64 & 70 \\
        Generator   & Duration  & 64 & 16 & 32 & 10 \\
                    & Rest      & 32 & 8  & 16 & 5 \\
        \midrule
        \multicolumn{2}{c|}{Discriminator} & 128+64+32 & 32 & 64 & logits \\
        \bottomrule
      \end{tabular}%
\end{table}

In ConMemofu and ConL2M models which exploit the RSE technique, the RSE vectors of music attributes are encoded by k-bin discritizers. The dimension of RSE for range feature is computed by $ k_{rng}^{\mathrm{attr}}=max(rng^{\mathrm{attr}}) - min(rng^{\mathrm{attr}})$. The dimension of pitch RSE for the average feature is computed by $ k_{avg}^{p}=round(max(avg^{\mathrm{attr}}) - min(avg^{\mathrm{attr}}))$. The dimension of duration and rest RSE for the average feature is designated to 10 because the class numbers of these two attributes are much fewer. The dimension of pitch, duration, and rest RSE for variance feature is designated to 30, 20, and 20, respectively.

In the adversarial steps of SeqMemofu and ConL2M, SeqLoss of each music attribute is exploited to update the parameters of the corresponding generator branch in a multi-task learning manner. 

\subsection{Objective evaluation}

To evaluate the proposed model, we generate melodies from the 805 sequences of lyrics in the test set and extract the music properties following \citep{yu_conditional_2021}. The detailed explanations of the measurements can be found in \citep{yu_conditional_2021}. The evaluation results of our proposed model against state-of-the-art models as baselines and other ablation models are shown in Table \ref{tab:metrics}.

% \begin{table}[htbp]\small
%     \centering
%     \begin{tabular}{lrrrr}
%     \hline
%     Metrics & \makecell[c]{Baseline} & \makecell[c]{Seq\\Melody} & \makecell[c]{SeqCtrl\\Melody} & \makecell[c]{GT} \\
%     \hline
%     MIDI span           & 12.05 & 10.72 & \textbf{11.27} & 12.05 \\
%     2-MIDI reps  & 5.45  & 5.12  & \textbf{5.84}  & 7.33  \\
%     3-MIDI reps  & 2.03  & 1.78  & \textbf{2.23}  & 3.72  \\
%     Unique MIDI & 0.58  & \textbf{0.52}  & 0.51  & 0.54  \\
%     Notes wo/ rest  & \textbf{15.91} & 16.01 & 16.00 & 15.82 \\
%     Average rest   & 0.58  & \textbf{0.52}  & 0.51  & 0.54  \\
%     Melody length       & 38.07 & 35.77 & \textbf{36.92} & 37.39 \\
%     \hline
%     \end{tabular}
%     \caption{Evaluation of proposed model against baseline.}
%     \label{tab:metrics}
% \end{table}

\begin{table}[htbp]
    \centering
    \begin{tabular}{lccccccc}
    \toprule
    Models & \makecell[c]{MIDI\\Span} & \makecell[c]{2-MIDI\\Reps} & \makecell[c]{3-MIDI\\Reps} & \makecell[c]{Unique\\MIDI} & \makecell[c]{Restless\\Notes}& \makecell[c]{Avg\\Rest} & \makecell[c]{Song\\Length} \\
    \midrule
    Ground-Truth  & 10.38 & 7.33 & 3.72 & 5.91 & 15.82 & 0.54 & 37.39 \\
    TBC-LSTM-GAN\cite{srivastava_melody_2022} & 12.05 & 5.45 & 2.03 & 6.97 & 15.91 & 0.58 & 38.07 \\
    C-Hybrid-GAN\cite{yu_conditional_2022} & 10.77 & 6.36 & 2.62 & 6.25 & 16.45 & 0.46 & 35.20 \\ 
    % Interpretable-GAN & 12.48 & 5.66 & 2.18 & 6.81 & 16.26 & 0.47 & 36.32 \\ 
    Memofu        & 11.72 & 5.84 & 2.27 & 6.71 & 16.19 & 0.52 & 37.02 \\
    SeqMemofu     & 11.98 & 5.22 & 1.87 & 7.06 & 16.62 & 0.45 & 34.49 \\
    ConMemofu     & 11.76 & 5.64 & 2.15 & 6.84 & 15.95 & 0.47 & 35.86 \\
    ConL2M (Proposed) & 10.75 & 5.86 & 2.23 & 6.61 & 15.61 & 0.52 & 37.32 \\
    \bottomrule
    \end{tabular}
    \caption{Evaluation of the proposed model against comparing models.}
    \label{tab:metrics}
\end{table}

Our proposed model outperforms other methods in most metrics. C-Hybrid-GAN better learns the repetitions in the pitch sequences thanks to the self-attention in the RMC. Overall, the proposed ConL2M achieves the best performance in objective evaluation.

\subsection{Style feature analysis}

\begin{figure}[htbp]
    \centering
    \includegraphics[width=0.9\columnwidth]{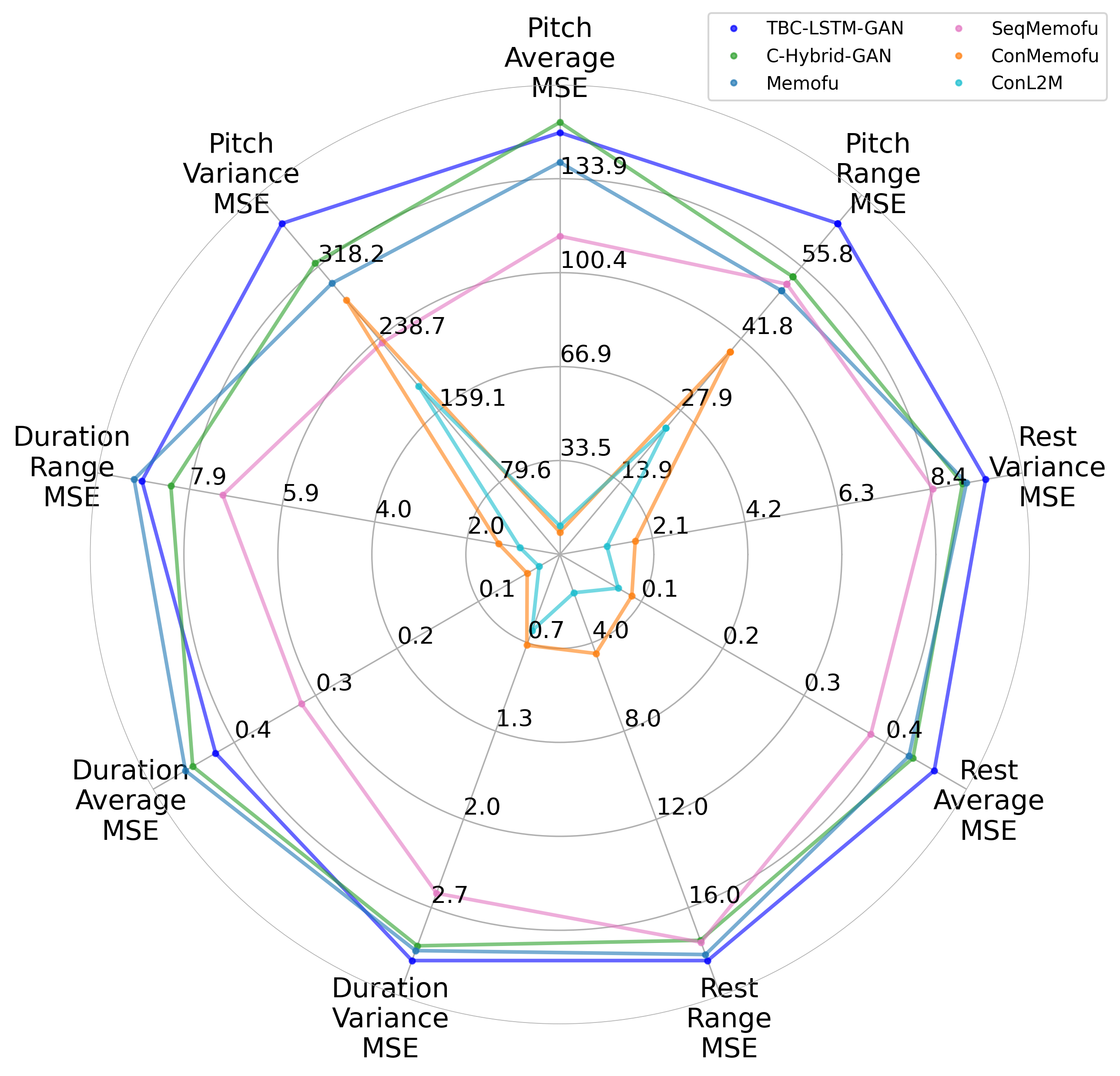}
    \caption{Radar plot of MSEs of attribute features.}
    \label{fig:radar}
\end{figure}

In order to evaluate the melody sequences generated by comparing models, we analyze the selected features of the music attributes of the generated melodies. For each pair of generated melody and ground-truth sequences, range, average, and variance of three attributes (pitch, duration, and rest) are calculated. Then, the MSEs on the test set are calculated. The evaluation results of different models are visualized by Fig. \ref{fig:radar}. 

Because the points on the axes represent the MSE of the corresponding feature, the generated quality is better when the occupied area is smaller on the radar chart. We can observe that the model with the proposed Memofu slightly outperformed the TBC-LSTM-GAN and C-Hybrid-GAN baselines. Moreover, adding the proposed SeqLoss further improves the performance. When the proposed RSE layer is introduced to the model, the generation accuracy has a significant improvement. This is also in accordance with our expectations because RSE offers the model more musical information for reference. Finally, our proposed ConL2M model achieves the best performance in the feature analysis experiment.

\subsection{Generation diversity analysis}

\begin{figure}[htbp]
    \centering
    \includegraphics[width=0.9\columnwidth]{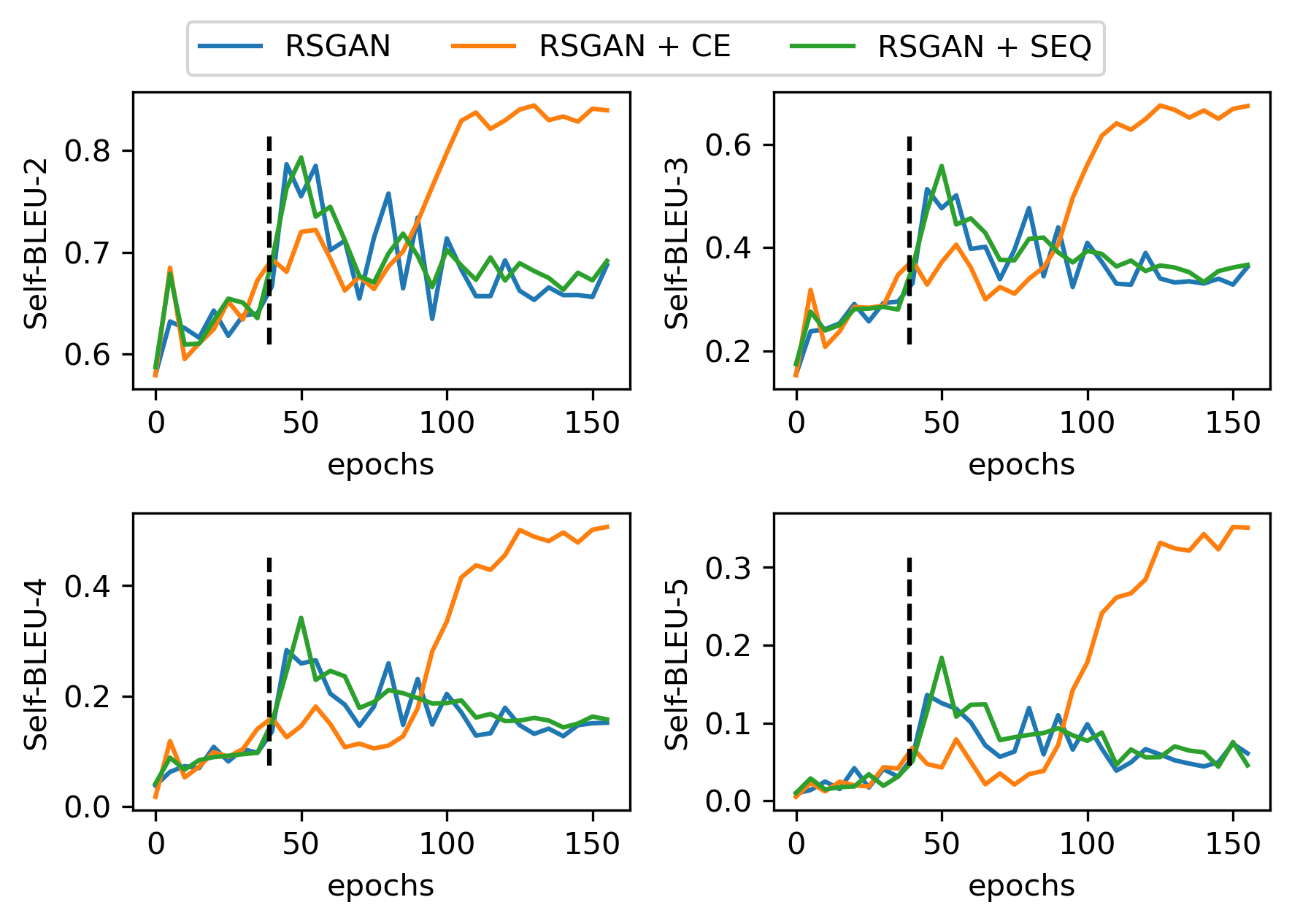}
    \caption{Generation diversity analysis of different loss functions.}
    \label{fig:bleu}
\end{figure}

Generation diversity is a very important metric in music generation tasks. A well-trained generative model should be able to generate diverse while high-quality content, which is much affected by the loss functions in the training procedure. In our work, the SeqLoss is proposed to improve the quality of generated melodies while preserving the generation diversity of GAN. To study the effects of different loss functions, we conduct an ablation study by training our model with i) RSGAN loss only, ii) RSGAN together with CE loss, and iii) RSGAN together with the proposed SeqLoss. During the training process, we record the Self-BLEU scores of the generated melodies in the validation dataset. The results are shown in Fig. \ref{fig:bleu}.

Self-BLEU score is used as a metric to measure the diversity of melodies generated by our proposed model. The value of the Self-BLEU score ranges between 0 and 1 with a smaller value of Self-BLEU implying a higher sample diversity and hence less chance of mode collapse in the GAN model. Intuitively, the Self-BLEU score measures how a generated melody sample is similar to the rest of the generated melody samples. 

\subsection{Controllability evaluation} \label{subsec:controllability}
In the training stage of ConL2M, RSEs from ground-truth data are fed into the model to offer style information of melodies. In the inference stage, thus, we can offer the RSEs of the desired musical style with input lyrics to control the generation process. The input RSE in the controllable generation process can be defined as follows:

\begin{equation} \label{eq:input_rse}
    \mathbf{RSE}^{\mathrm{attr}}_{in} = [ f_p({rng}^{\mathrm{attr}}_{norm}), f_d({avg}^{\mathrm{attr}}_{norm}), f_r({var}^{\mathrm{attr}}_{norm})], 
\end{equation}
where $rng_{norm}$, $avg_{norm}$, and $var_{norm}$ are normalized values in $[0,1]$ of the desired range, average, and variance, respectively. $f(\cdot)$ denotes an embedding function that converts the normalized feature values to the RSE vectors. First, a linear mapping function reverts the normalized values to the real values in the dataset. Then, the feature values are embedded as one-hot vectors using the same k-bin discretizers of the training stage.

\begin{figure}
     \centering
     \begin{subfigure}[b]{\columnwidth}
         \centering
         \includegraphics[width=0.9\columnwidth]{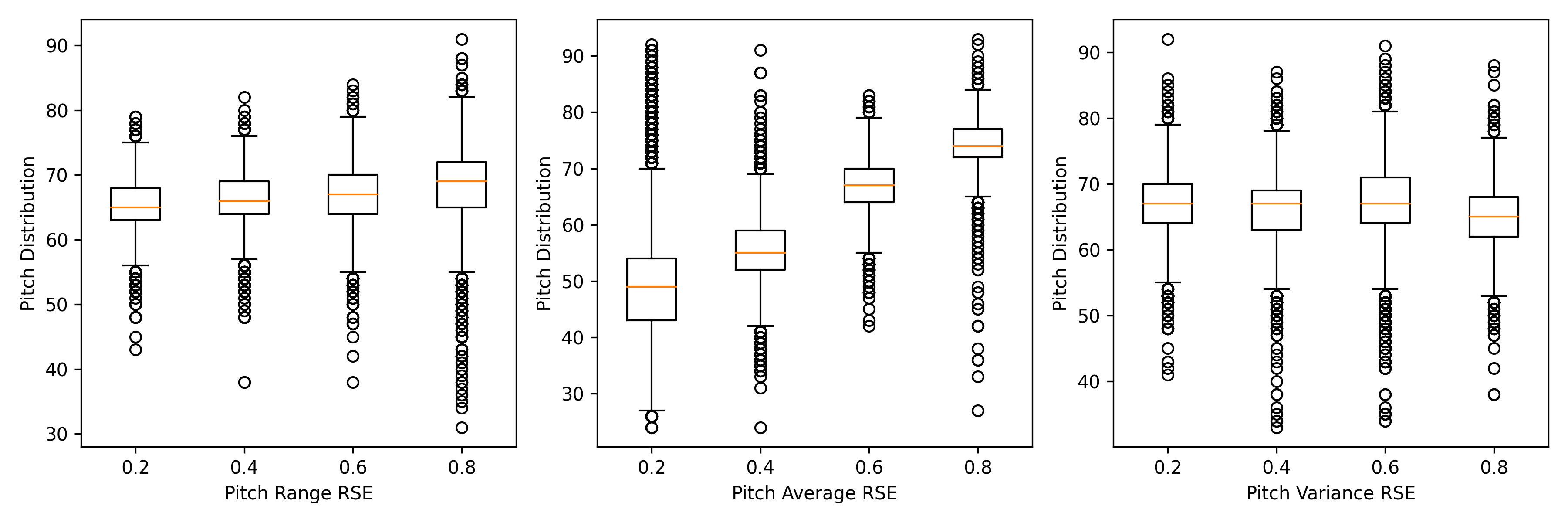}
         \caption{Distributions of pitch values in the generated sequences given different input values of pitch control RSEs.}
         \label{fig:controllability_pitch}
     \end{subfigure}
     \begin{subfigure}[b]{\columnwidth}
         \centering
         \includegraphics[width=0.9\columnwidth]{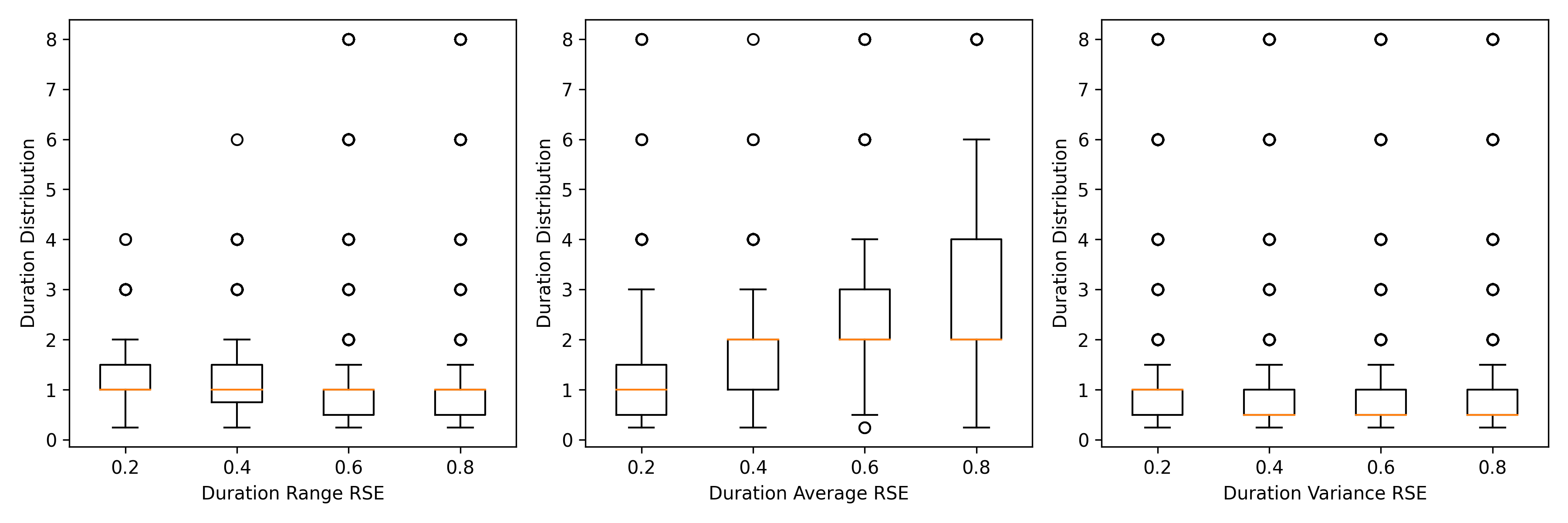}
         \caption{Distributions of duration values in the generated sequences given different input values of duration control RSEs.}
         \label{fig:controllability_duration}
     \end{subfigure}
     \begin{subfigure}[b]{\columnwidth}
         \centering
         \includegraphics[width=0.9\columnwidth]{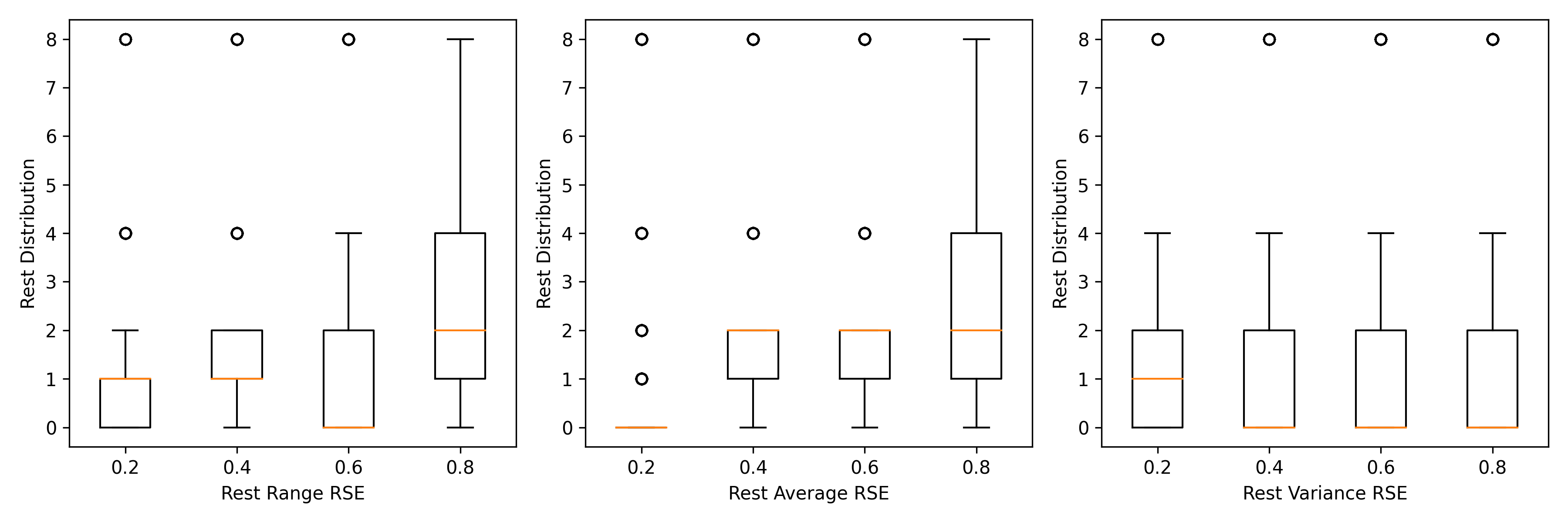}
         \caption{Distributions of rest values in the generated sequences given different input values of rest control RSEs.}
         \label{fig:controllability_rest}
     \end{subfigure}
        \caption{Controllability evaluation experiments.}
        \label{fig:controllability_expriment}
\end{figure}

In order to quantitatively evaluate the controllability introduced by the proposed RSE technique, we choose a set of values as the candidates, $[0.2, 0.4, 0.6, 0.8]$, as the normalized value features. For all the melody features, namely range, average, and variance of pitch, duration, and rest, we repeat the control variable experiment of lyrics-to-melody generation. In each experiment, the controlled RSE input iterates in the candidate set while other RSE inputs remain the same constants. Then, every RSE input is used to offer control signals for all the melodies in the test dataset. The distribution of the corresponding music attributes in all generated melody sequences is visualized by the boxplots, as shown in Fig. \ref{fig:controllability_expriment}.

From the music attribute distribution given different RSE inputs, we can see that the generated melodies are effectively controlled by the RSE inputs. When increasing the RSEs of range of the music attributes, the generated sequences obviously have a larger range of distribution. When increasing the RSEs of average of the music attributes, the generated sequences tend to distribute in higher value range. For the RSEs of variance, it is not easy to observe the patterns in the boxplots due to the nature of boxplot visualization, but the controllability of variance can be verified by other evaluation methods in this section. Overall, the proposed RSE technique demonstrates significant controllability of the lyric-to-melody generation process in the experiment.

\subsection{Sheet music case study} \label{subsec:sheetmusic}

To study the controllability provided by RSE, we generate sheet music with different models and input RSE from the same lyrics. In the inference, we represent input RSE by triplets of features as ${RSE}^{\mathrm{attr}}_{in} = [{rng}^{\mathrm{attr}}, {avg}^{\mathrm{attr}}, {var}^{\mathrm{attr}}]$, where $rng$, $avg$, and $var$ are normalized values in $[0,1]$ of the desired range, average, and variance, respectively.

% trim={<left> <lower> <right> <upper>}
\begin{figure}
     \centering
     \begin{subfigure}[b]{0.9\columnwidth}
         \centering
         \includegraphics[width=\columnwidth, trim=50 720 50 55]{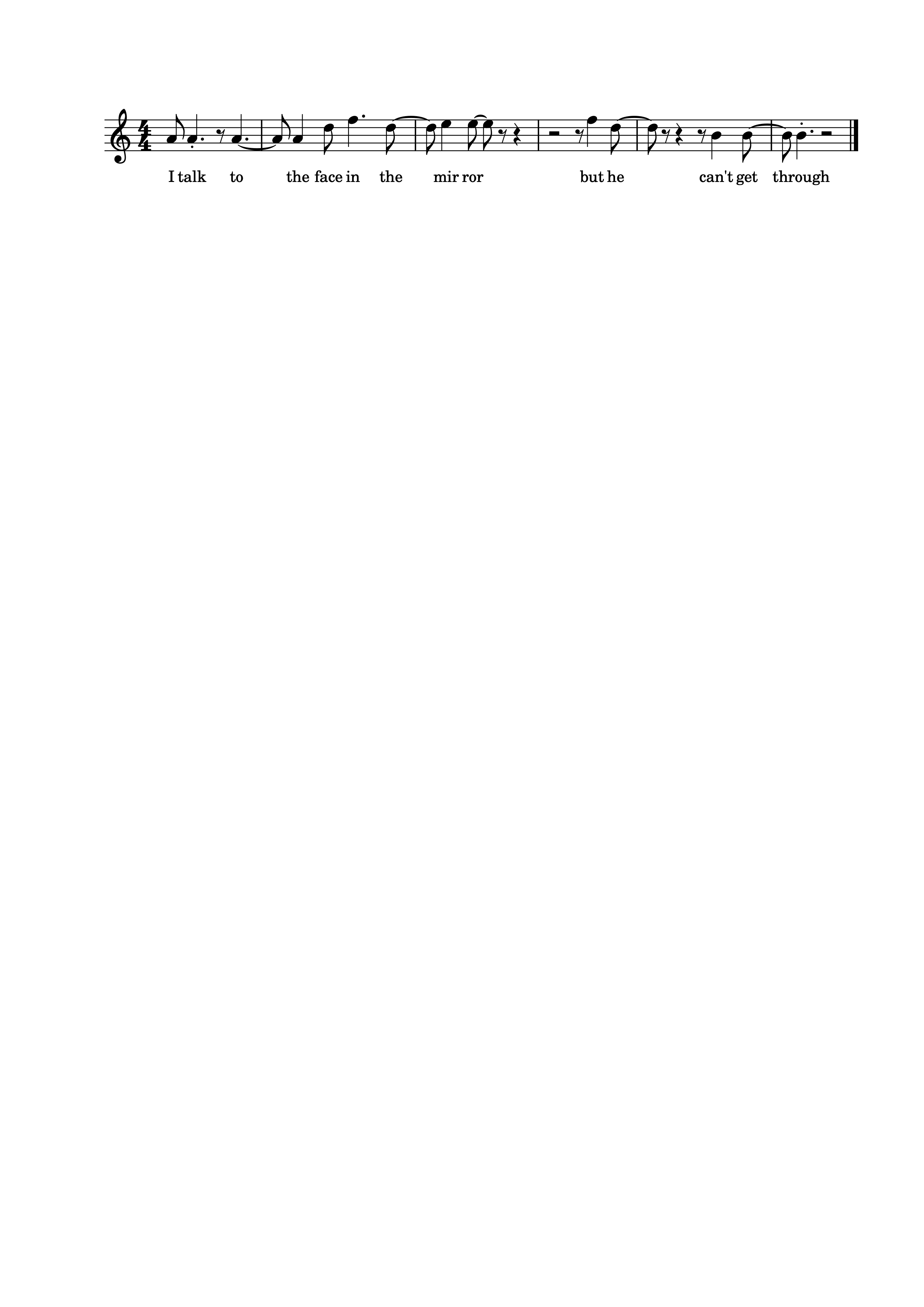}
         \caption{Generated by TBC-LSTM-GAN.}
         \label{fig:sheet_baseline}
     \end{subfigure}
     % \hfill
        \begin{subfigure}[b]{0.9\columnwidth}
         \centering
         \includegraphics[width=\columnwidth, trim=50 720 50 55]{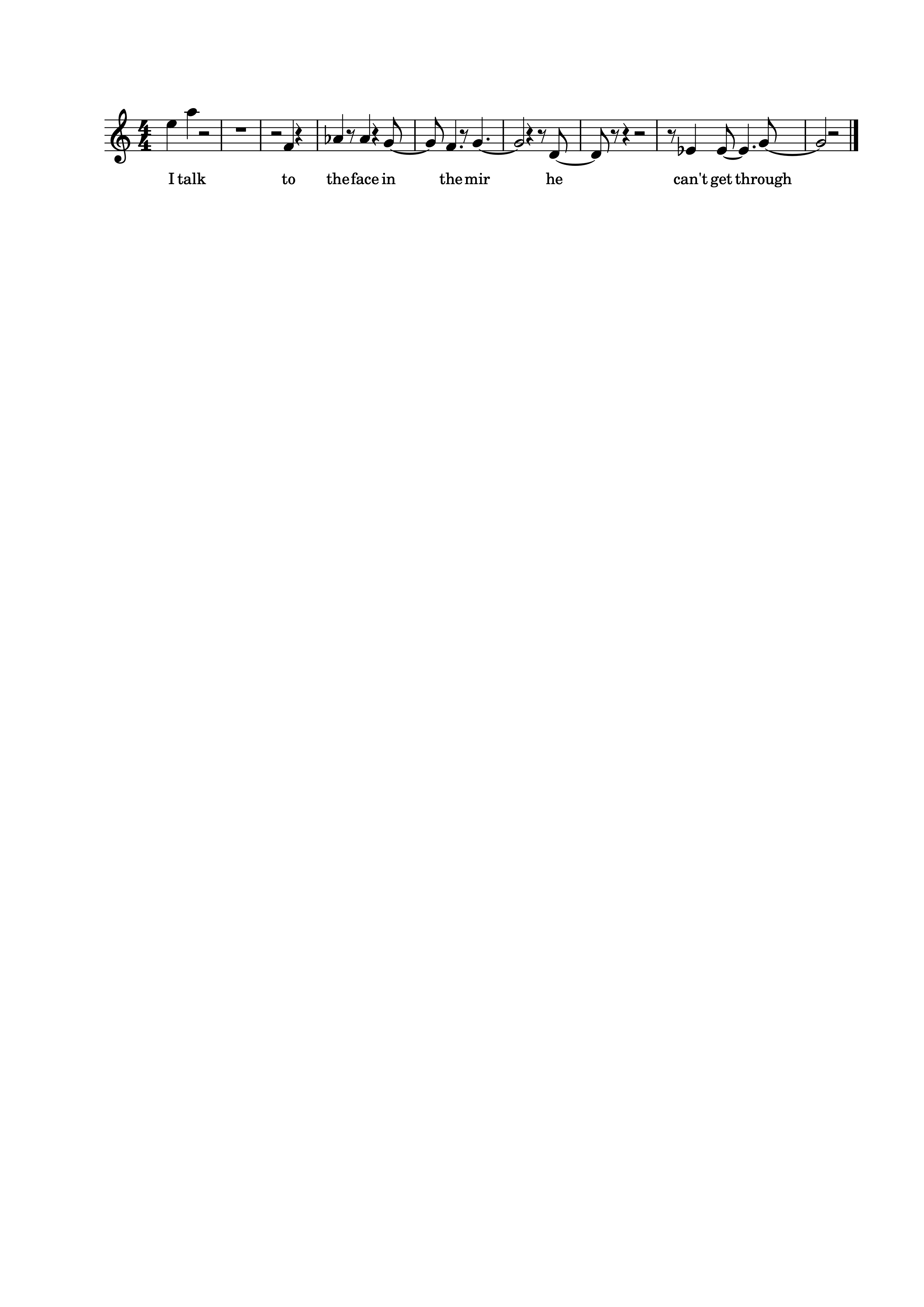}
         \caption{Generated by C-Hybrid-GAN.}
         \label{fig:sheet_hybrid}
     \end{subfigure}
     % \hfill
     \begin{subfigure}[b]{0.9\columnwidth}
         \centering
         \includegraphics[width=\columnwidth, trim=50 720 50 55]{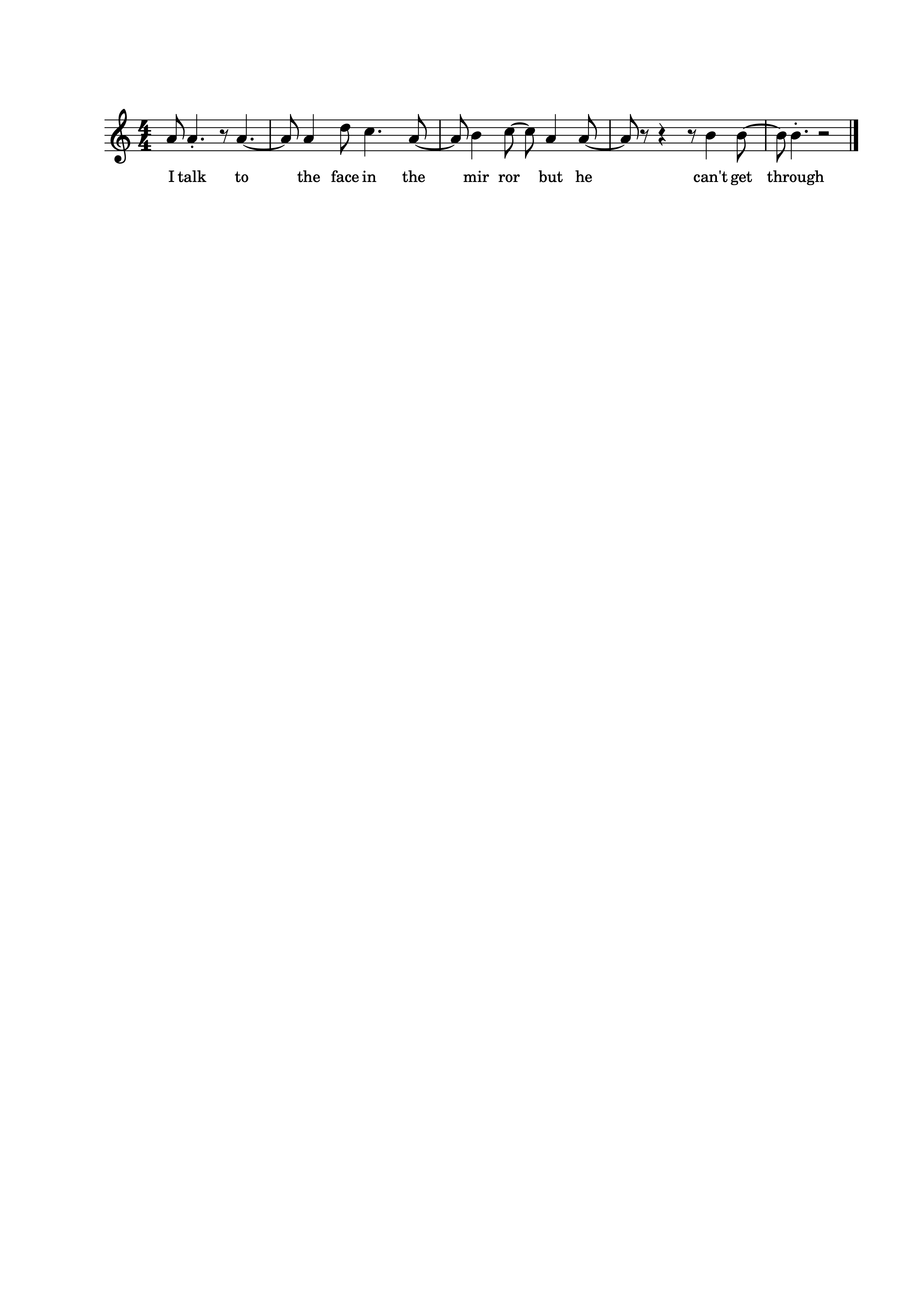}
         \caption{Generated by SeqMemofu.}
         \label{fig:sheet_seq}
     \end{subfigure}
     % \hfill
     \begin{subfigure}[b]{0.9\columnwidth}
         \centering
         \includegraphics[width=\columnwidth, trim=50 720 50 55]{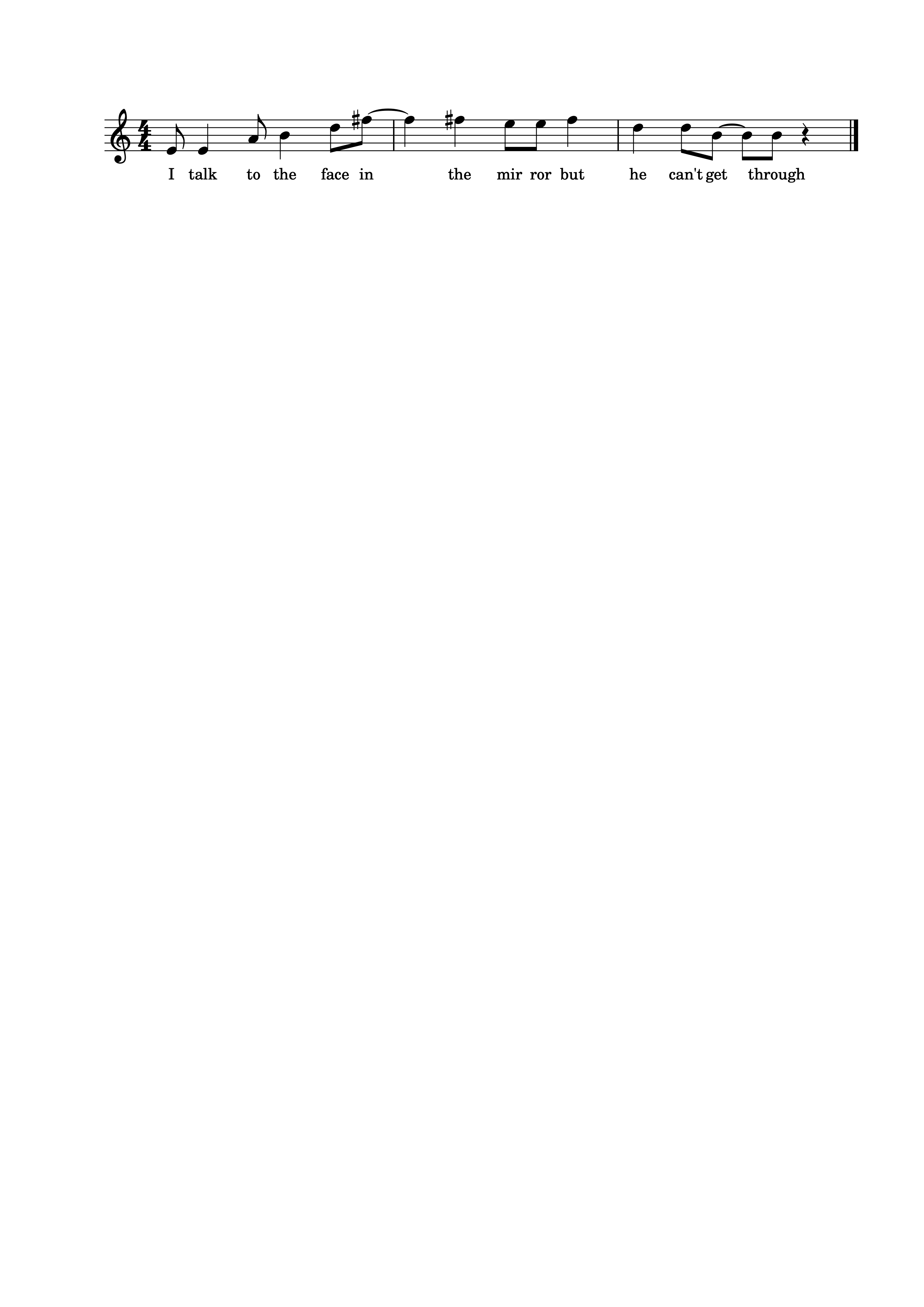}
         \caption{${RSE}^{p}_{in}$=[0.7,0.7,0.8], ${RSE}^{d}_{in}$=[0.1,0.1,0.1], and ${RSE}^{r}_{in}$=[0.1,0.1,0.1].}
         \label{fig:sheet_styleseq0}
     \end{subfigure}
     % \hfill
     \begin{subfigure}[b]{0.9\columnwidth}
         \centering
         \includegraphics[width=\columnwidth, trim=50 720 50 55]{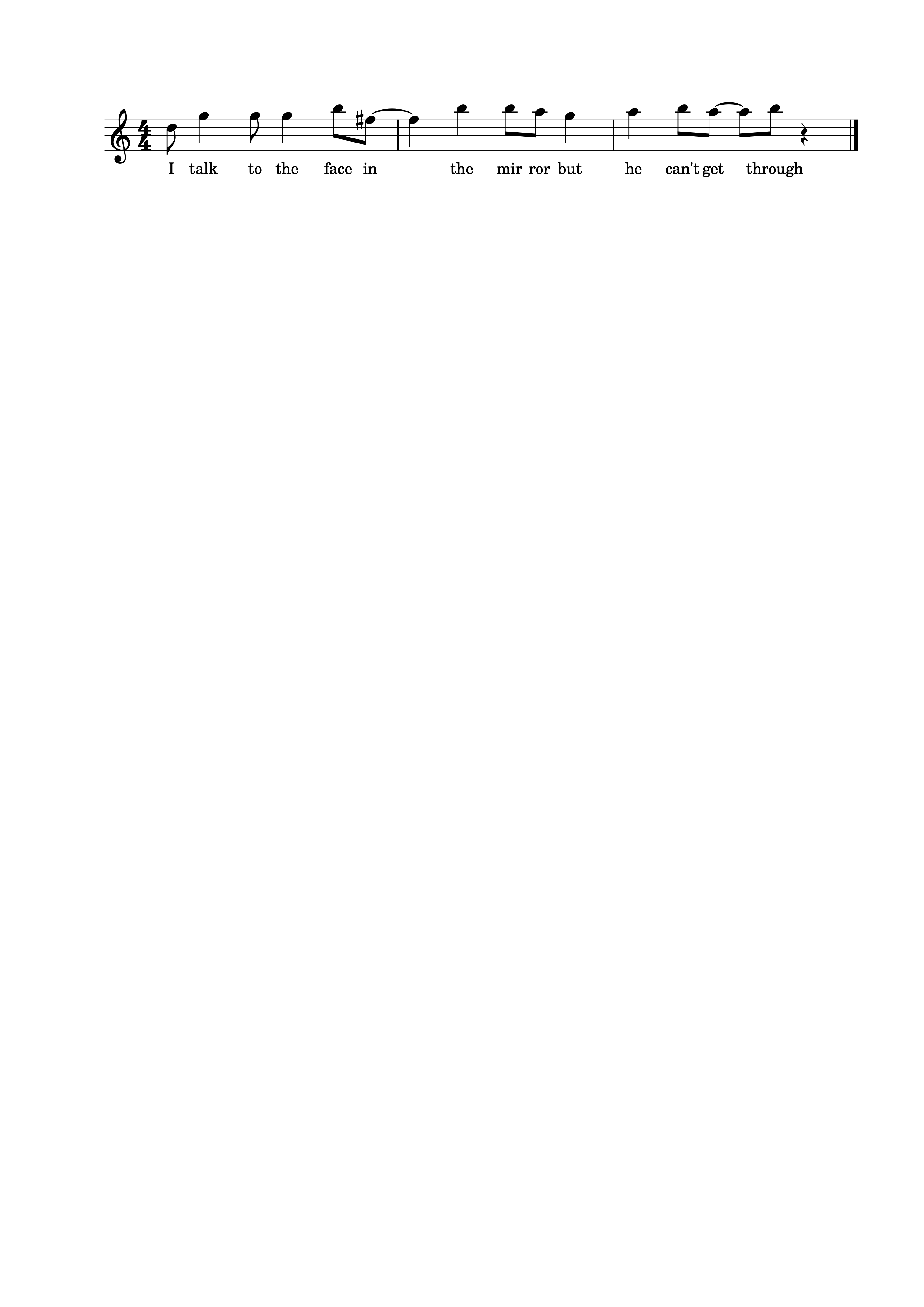}
         \caption{${RSE}^{p}_{in}$=[0.3,0.9,0.1], ${RSE}^{d}_{in}$=[0.1,0.1,0.1], and ${RSE}^{r}_{in}$=[0.1,0.1,0.1].}
         \label{fig:sheet_styleseq1}
     \end{subfigure}
     % \hfill
     \begin{subfigure}[b]{0.9\columnwidth}
         \centering
         \includegraphics[width=\columnwidth, trim=50 720 50 55]{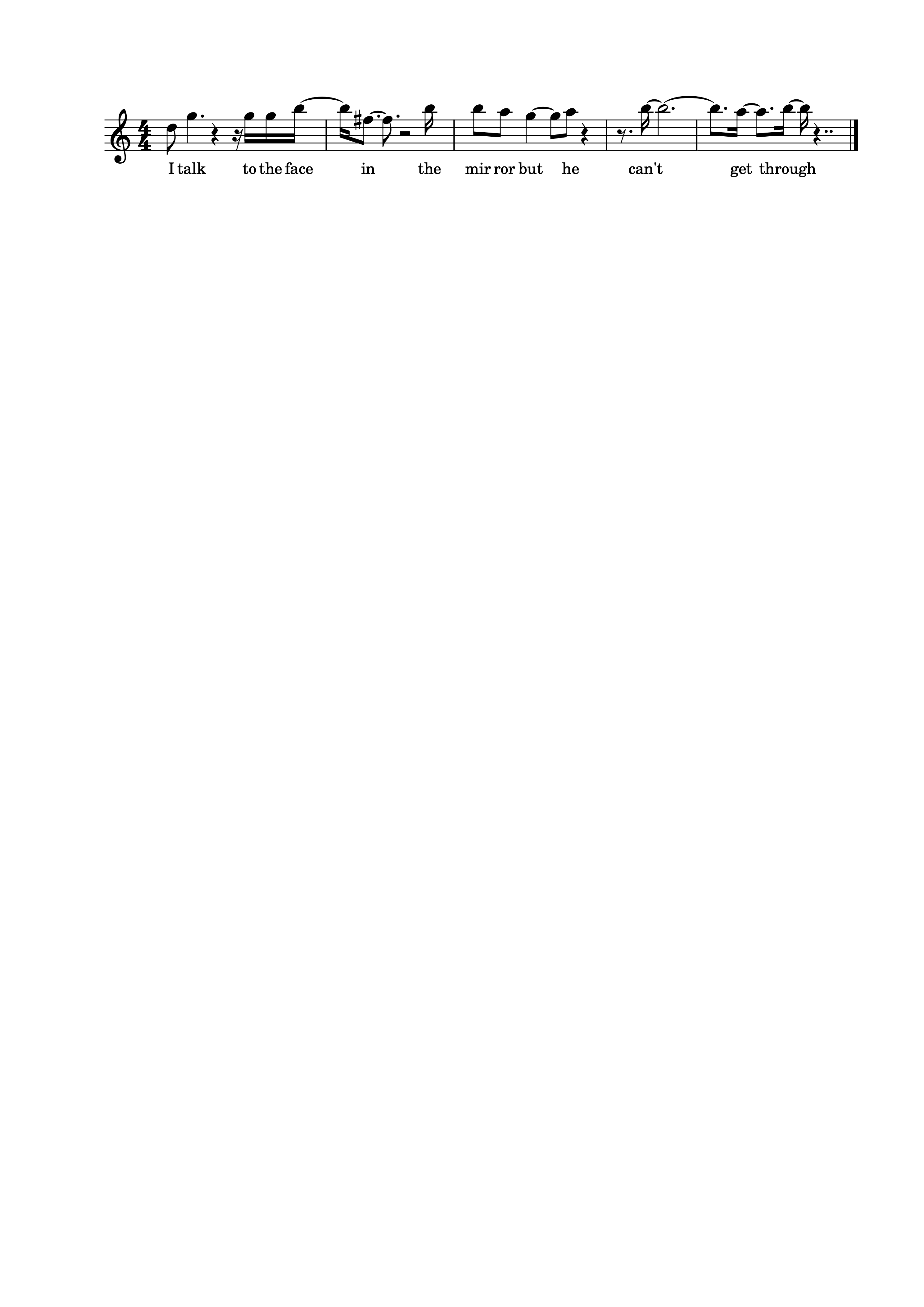}
         \caption{${RSE}^{p}_{in}$=[0.3,0.9,0.1], ${RSE}^{d}_{in}$=[0.2,0.1,0.3], and ${RSE}^{r}_{in}$=[0.3,0.1,0.2].}
         \label{fig:sheet_styleseq2}
     \end{subfigure}
        \caption{Generated sheet music.}
        \label{fig:sheetmusic}
\end{figure}

Fig. \ref{fig:sheetmusic} shows an example of generated sheet music given lyrics. Comparing (a), (b), and (c), we can see SeqMemofu generates more smooth melody contour than TBC-LSTM-GAN and C-Hybrid-GAN and manages to avoid unreal rest notes. In (d), by randomly sampling a set of RSE inputs, Memofu shows the ability of capturing the rhythm of lyrics. Then, we modify the RSE with higher pitch average in (e), and ConL2M generates notes with higher pitch. Moreover, by modifying RSE of duration and rest with higher range and variance in (f), ConL2M generates complex rhythm of melody with the same pitch contour and lyrics. Through analysis and comparison of the generated sheet music, our proposed ConL2M not only generates better melodies, but also demonstrates significant controllability consistent with human perception of music.

\section{Subjective Evaluation}

It is known that music evaluation is a very subjective recognition process related to personality, background, and taste. Therefore, on top of objective evaluation, we also conduct subjective evaluations to study the generation quality and controllability of the proposed models. In our subjective evaluation section, seven volunteers with music knowledge at different levels are involved as subjects. All the results are submitted anonymously without personal information in the following evaluation experiments. 
% The survey and generated samples can be accessed via Google Form\footnote{\url{https://forms.gle/iC4cdtJcvUXs9VCc7}}.

\subsection{Generation quality evaluation}

\begin{figure}[htbp]
    \centering
    \includegraphics[width=0.9\columnwidth]{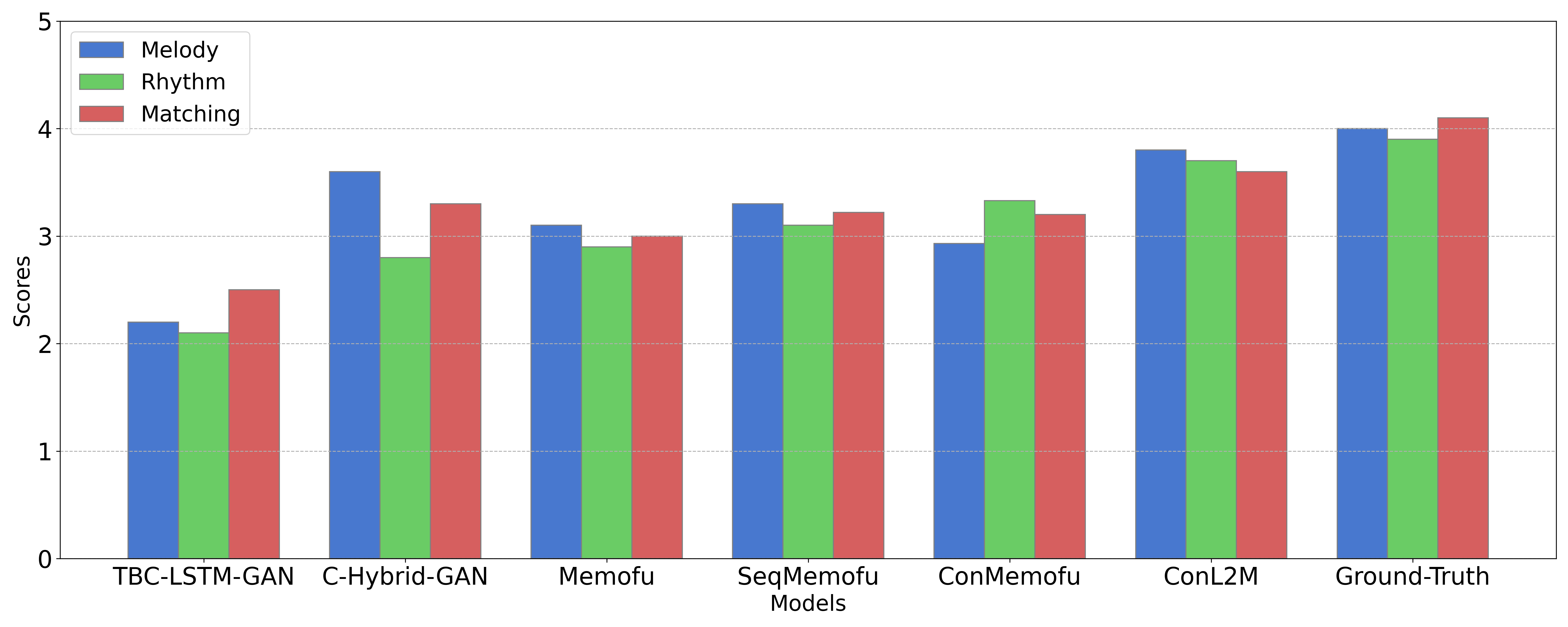}
    \caption{Results of subjective evaluation of generation quality.}
    \label{fig:subjective_quality}
\end{figure}

In the subjective evaluation for generation quality, we generate melodies using the above five competing models given three different pieces of input lyrics. Then singing voice audio is synthesized by Synthesizer V\footnote{\url{https://dreamtonics.com/synthesizerv/}}, a singing voice synthesizing software, from generated melodies in MIDI format and the corresponding lyrics. Thus, 18 singing voice audio of generated melodies are prepared to be rated by the subjects. We release and distribute the audio files for the subjects and the subjects can locally play and listen to the audio. The sheet music of the melodies is also provided with the audio files as a supplementary for those subjects who hope to read. In particular, each melody is played three times in random order. After listening to each melody three times, we asked the subjects three questions as subjective evaluation metrics for melody quality:
\begin{itemize}
    \item How good do you feel about the melody? 
    \item How good do you feel about the rhythm?
    \item How fitting do you feel about the melody and the lyrics?
\end{itemize}
Then, the subjects give a score from 1 to 5 (very bad, bad, okay, good, very good) to rate the generated melody for each question. The averaged evaluation results are shown in Fig. \ref{fig:subjective_quality}.

Evaluation results show that our proposed ConL2M performs best through all the comparing models, approaching the ground-truth. Meanwhile, there is still a gap between generated melodies and ground-truth ones, indicating the potential of further improving the generation quality.

\subsection{Controllability evaluation}

Controllability of the melody generation process is an important aspect of our work. Therefore, we propose a subjective evaluation method and design experiment to demonstrate the controllability of the proposed ConL2M model. Firstly, we choose four aspects of musical style as the feature that we desire to control, which are i) pitch distribution, ii) melody contour, iii) tempo speed, and iv) rhythm pattern. For each style feature to be studied, we generate melodies from lyrics with three different sets of RSE inputs. The values of the RSE inputs are designated to generate three different melodies and corresponding audio in the desired musical style. Then, for each feature group of generated melodies, we ask the subjects to rank the melodies according to the specific feature after listening in random order. Specifically, the ranking questions for each style feature are listed below:
\begin{itemize}
    \item Pitch distribution: please rank the pitch of the melody from low to high.
    \item Melody contour: please rank the melody contour from flat to fluctuate.
    \item Tempo speed: please rank the tempo speed from slow to fast.
    \item Rhythm pattern: please rank the rhythm pattern from steady to complex.
\end{itemize}
If the answer of the subject is the same order as the input RSE intends to generate, we get 1 score. Or else, we get 0. In such way, we use 5 pieces of different lyrics to generate melodies to be evaluated and accumulate the scores for the controlling style features. Finally, we get a 5-scale score for each style feature after averaging the subjects. The evaluation results are shown in Table \ref{tab:subjective_controllability}.

\begin{table}[htbp]
    \centering
    \begin{tabular}{lcccc}
    \toprule
    \makecell[c]{Control\\Feature} & \makecell[c]{Pitch\\Distribution} & \makecell[c]{Melody\\Contour} & \makecell[c]{Tempo\\Speed} & \makecell[c]{Rhythm\\Pattern} \\
    \midrule
    Score & 4.92 & 3.63 & 4.76 & 3.52 \\
    \bottomrule
    \end{tabular}
    \caption{Subjective evaluation of style controllability.}
    \label{tab:subjective_controllability}
\end{table}
Among the tests, the subjects get a high score in ranking pitch distribution and tempo speed, proving significant controllability of the proposed ConL2M. Relative lower scores for recognizing melody contour and rhythm pattern may partly owe to the indirectness of such musical features. The results of controllability experiments also inspire us to design higher-level and more explicit human interaction for controllable music generation in future work.

\section{Conclusion}

To address the generation quality and controllability issues in the lyrics-to-melody generation task, we proposed ConL2M in this paper towards controllable lyrics-to-melody generation. SeqLoss is exploited to help the model better learn the correlation between lyrics and melody data and meaningful information at the sequence level. RSE technique realizes significant controllability of the generation process as well as improves generation quality. Experiment results and evaluation metrics confirm that ConL2M has great power of composing melody given lyrics as an AI songwriter. Moreover, our proposed techniques can be conveniently extended to some suitable tasks, such as symbolic music generation conditioning on input sequences or labels. Future work will focus on data augmentation methods for the lyrics-melody dataset, advanced generation architecture, and better human interaction pipelines for controlling the generation process.

\bibliographystyle{unsrtnat}
\bibliography{references}  %%% Uncomment this line and comment out the ``thebibliography'' section below to use the external .bib file (using bibtex) .





% \end{thebibliography}

\end{document}